\documentclass[12pt]{article}
\usepackage{amsmath,amssymb,exscale}
\usepackage{color,graphicx,epsfig}
\usepackage[utf8]{inputenc}
\usepackage{hyperref}
\input epsf
\newcommand{\pslash}{{\not \!p}}

\newcommand{\m}[1]{\marginpar{{\tiny *}} }

\def\be{\begin{equation}}
\def\te{\end{equation}}

\def\0{{(0)}}

\def\al#1\fal{\begin{align}#1\end{align}}

\def\dg{\dagger}
\def\su{\tilde{R}_2}
\def\sd{\hat{R}_2}
\def\suh{\tilde{R}_2^\dagger}
\def\sdh{\hat{R}_2^\dg}
\def\hc{\textrm{h.c.}}

\begin{document}
\topmargin -1.0cm
\oddsidemargin 0cm
\evensidemargin 0cm

\thispagestyle{empty}

\vspace{40pt}

\begin{center}
\vspace{40pt}

\Large \textbf{Composite Higgs and leptoquarks from a simple group}

\end{center}

\vspace{15pt}
\begin{center}
{\bf Leandro Da Rold$^{\dagger}$, Federico Lamagna$^{\star}$} 

\vspace{20pt}

\textit{Centro At\'omico Bariloche, Instituto Balseiro and CONICET}
\\[0.2cm]
\textit{Av.\ Bustillo 9500, 8400, S.\ C.\ de Bariloche, Argentina}

\end{center}

\vspace{20pt}
\begin{center}
\textbf{Abstract}
\end{center}
\vspace{5pt} {\small \noindent
We propose a composite grand unified theory to study the anomalies in the semileptonic $B$ decays. 
We show a simple group containing the custodial and Standard Model gauge symmetries, that can deliver a set of composite pseudo Nambu-Goldstone bosons: the Higgs, a colorless SU(2)$_L$-fourplet and three leptoquarks: a triplet and two doublets. We give a description in terms of an effective theory of resonances.
By assuming anarchic partial compositeness of the Standard Model fermions, we find representations for the composite fermions that allow to obtain the Higgs Yukawa couplings, as well as leptoquark interactions explaining the deviations in $R^{\mu e}_{K^{(*)}}$.
We calculate the one-loop potential, we show that it can trigger electroweak symmetry breaking and we find a region of the parameter space that can reproduce the Standard Model spectrum. 
The model predicts  leptoquark masses of order $0.4-1.3$~TeV, corrections to some electroweak observables, with $Zb_L\bar b_L$ saturating the current bounds, and a very reach phenomenology at LHC.
We also study the possibility of explaining $R^{\tau\ell}_{D^{(*)}}$.
}

\vfill
\noindent {\footnotesize E-mail:
$\dagger$ daroldl@cab.cnea.gov.ar,
$\star$ federico.lamagna@cab.cnea.gov.ar
}

\noindent
\eject

%\tableofcontents

\section{Introduction}
The experiments at LHC have discovered the Higgs, measuring many of its properties. Despite the impressive level of accuracy of these measurements, the question about the nature of the Higgs and the dynamics that stabilises its potential is still open, since neither new particles, nor large deviations associated to Higgs physics, have been found yet. 

Besides the Higgs discovery, in the recent years different experiments have provided very interesting results related with flavor physics: a set of anomalies in semileptonic decays of $B$-mesons, that could be a hint of violation of lepton flavor universality. The ratios $R^{\mu e}_{K^{(*)}}$~\cite{Aaij:2017vbb} and $R^{\tau\ell}_{D^{(*)}}$~\cite{Lees:2013uzd,Hirose:2016wfn,Aaij:2015yra} show deviations with respect to the Standard Model (SM) of order $4\sigma$, associated to transitions of quarks in neutral and charged currents, respectively. 
These deviations could be explained introducing New Physics (NP) at the scale of a few TeV, mainly coupled with the third generation. Recent studies have shown that the presence of a vector leptoquark transforming as $({\bf 3},{\bf 1})_{2/3}$ under the SM gauge symmetry group~\cite{Assad:2017iib,Blanke:2018sro,Fornal:2018dqn}, or the presence of two scalar leptoquarks, transforming as $(\bar{\bf 3},{\bf 1})_{1/3}$ and $(\bar{\bf 3},{\bf 3})_{1/3}$, can explain the deviations in both observables, without introducing incompatibilities in other low-energy observables, or in high-energy searches at LHC~\cite{Buttazzo:2017ixm,Crivellin:2017zlb}. Ref.~\cite{Becirevic:2018afm} has also considered scalars in a $({\bf 3},{\bf 2})_{7/6}$ and a $(\bar{\bf 3},{\bf 3})_{1/3}$.

A strongly coupled field theory (SCFT), with resonances at the scale of few TeV, could provide a solution to the stability of the Higgs potential, as well as to the $B$-anomalies. In this framework, partial compositeness of the SM fermions, not only gives a rationale for the flavor puzzle, but also explains the preferential coupling of the resonances with the third generation. To avoid a bunch of new states at the TeV scale, that could be in conflict with precision observables, it would desirable to introduce a gap in the SCFT spectrum, allowing a set of leptoquarks to be lighter than the rest of the resonances. In the case of scalar leptoquarks, this gap can be obtained if they are Nambu-Goldstone bosons (NGBs) of the SCFT dynamics.
If the interactions with the SM explicitly break the global symmetries of the SCFT, a potential for the leptoquarks is generated at loop-level. If the Higgs is also a pseudo Nambu Goldstone boson (pNGB), the same potential that produces leptoquark masses could be able to trigger electroweak (EW) symmetry breaking. The author of Ref.~\cite{Marzocca:2018wcf} has shown the fundamental description of an SCFT that, under some suitable assumptions, delivers the appropriate set of leptoquarks, as well as the Higgs. The lack of non-perturbative methods is a limitation for the quantitative predictions of this description. Ref.~\cite{Gripaios:2014tna} has provided an effective description of a model with a Higgs and leptoquarks as pNGBs, based on a factorizable group: SO(9)$\times$SO(5), spontaneously broken down to SO(6)$\times$SO(3)$\times$SO(4). The Higgs is in the usual coset of the Minimal Composite Higgs Model (MCHM)~\cite{Agashe:2004rs}: ${\rm SO}(5)/{\rm SO(4)}$, whereas the leptoquarks are in the coset ${\rm SO}(9)/[{\rm SO}(6)\times {\rm SO}(3)]$, that leads to a multiplet in the representation $(\bar{\bf 3},{\bf 3})_{1/3}$.

In the present work we embed the framework of Ref.~\cite{Gripaios:2014tna} in a simple group G, obtaining a composite grand unified theory. We show a pattern of spontaneous symmetry breaking that generates the Higgs and $S_3$ as NGBs, as well as a colorless SU(2)$_L$-fourplet and two new leptoquark doublets. We find embeddings of fermionic SCFT operators that, after mixing with the SM fermions, lead to the right Higgs Yukawa couplings, as well as to the leptoquark couplings required to explain $R_{K^{(*)}}$. The mixing generates a potential at one loop that can trigger EW symmetry breaking and generate leptoquark masses dynamically. Unification provides a highly predictive scenario, relating the Higgs and the leptoquark sectors.

In order to make precise predictions, we consider a description of the resonances of the SCFT in terms of a two-site theory, that provides a weakly coupled description of the composite dynamics. We compute the potential at one loop and show that there are regions of the parameter where EW symmetry breaking (EWSB) occurs dynamically. We also compute the masses of the would-be NGBs.

We discuss some issues related with EW precision tests, as the $\rho$ parameter and corrections to $Zb_L\bar b_L$. 
We also consider the possibility of explaining $R_{D^{(*)}}$ with the leptoquark content of the theory. We show that bounds from processes as: lepton flavor universality violation in $\tau$ decays, as well as $B\to K\nu\bar\nu$, are not compatible with $R_{D^{(*)}}$, in agreement with results from the literature~\cite{Buttazzo:2017ixm,Angelescu:2018tyl}. A solution to this puzzle could be generated by including a leptoquark $S_1\sim(\bar{\bf 3},{\bf 1})_{1/3}$, or $R_2\sim({\bf 3},{\bf 2})_{7/6}$. It is very simple to include scalar states with those charges, but as ordinary resonances, expected to be heavier than the NGBs. A model fully addressing the $B$-anomalies requires an extension of our model, including $S_1$ or $R_2$ as a NGBs, also. The present work represents a first step towards that solution, in terms of an effective theory of resonances. 

The paper is organised as follows:  in sec.~\ref{sec-model} we describe an SCFT based on symmetry principles. We discuss its global symmetry group and the pattern of spontaneous symmetry breaking leading to NGBs, containing the Higgs and $S_3$. We also select representations of the fermionic operators of the SCFT that allow to obtain suitable Yukawa couplings. We briefly discuss some properties of the global symmetries related with physics constraints. In sec.~\ref{eff-th} we describe the effective theory obtained after integration of the heavy states of the model, that contains the SM degrees of freedom and the NGBs. This effective theory allows us to compute the potential, and study the conditions that lead to an appropriate vacuum, this is shown in sec.~\ref{sec-potential}. We study the phenomenology of the leptoquarks in sec.~\ref{sec-pheno} and we conclude in sec.~\ref{sec-conclusions}.

\section{Leptoquarks and Higgs as composite pNGBs}\label{sec-model}
The composite-Higgs model that we will consider has the following structure. There is a sector of elementary fields, containing the same degrees of freedom as the SM, except the Higgs, that is not present in this sector. There is also a new strongly interacting sector, that produces bound states, or resonances, at a scale $m_*$ of few TeV. The resonances interact with couplings collectively denoted as $g_*$, that will be assumed to be perturbative, in the range: $g_{\rm SM}\lesssim g_*<4\pi$. This sector has a global symmetry G, with G a simple group, that contains the SM gauge symmetry. G is spontaneously broken by the strong dynamics to a subgroup H, generating a set of NGBs that can be parametrised by the broken generators in G/H. We will focus here on the case where the set of NGBs contains, at least, the Higgs as well as a leptoquark $S_3$. The NGBs decay constant is of order: $f\sim \sqrt{2}m_*/g_*$. We assume that there are fermionic resonances transforming in irreducible representations of G. It is also straightforward to include spin-one resonances transforming with the adjoint representation of G (they can be excited by the Noether currents of the SCFT associated to the symmetry G). The SCFT sector will be taken flavor anarchic, thus all the Yukawa couplings of the fermion resonances are of the same order.

The elementary sector and the SCFT, or composite sector, interact with each other. The elementary gauge fields weakly gauge a subgroup of G. We assume that, at a high ultra-violet scale $\Lambda$, the elementary fermions have linear interactions with the SCFT: ${\cal L}_{\Lambda}\supset \omega\ \bar\psi {\cal O}^{\rm SCFT}$, with ${\cal O}^{\rm SCFT}$ fermionic operators. These operators, being defined at a scale $\Lambda\gg f$, transform linearly with irreducible representations of G. These representations are not fixed, leaving room for model-building, we will discuss the conditions they must satisfy, selecting a suitable set of them, in sec.~\ref{sec-frep}. 

At energies of order $m_*$, a linear mixing between the elementary fermions and the composite resonances $\Psi$, created by ${\cal O}^{\rm SCFT}$, is generated:
\begin{equation}\label{eq-pceff}
{\cal L} \supset f\lambda_\psi\ \bar\psi \Psi +{\rm h.c.} \ ,
\end{equation}
%where $\lambda\sim\omega(m_*)$. We will work in the basis where $\lambda$ are diagonal.
This mixing can be diagonalised by a simple rotation, leading to a partially composite massless state, with degree of compositeness:
\begin{equation}\label{eq-epsilon}
\epsilon_\psi\equiv\frac{\lambda_\psi}{g_*} \ .
\end{equation}
These states have Yukawa interactions with the NGBs of order: $y\sim \epsilon_L g_* \epsilon_R$. After EWSB these states become massive, with masses of order $yv/\sqrt{2}$.

Assuming that the evolution of ${\cal O}^{\rm SCFT}$ is driven by its scaling dimension $\Delta$, the coupling of Eq.~(\ref{eq-pceff}) is of order: $\lambda\sim\left(\frac{m_*}{\Lambda}\right)^{\Delta-5/2}$. As is well known, if $\Lambda\gg m_*$, hierarchically small couplings can be generated if $\Delta>5/2$.

Flavor can be introduced by adding generation indices to the elementary fermions and to the SCFT operators. In this case the couplings $\lambda$ become tensors in flavor space, that can be diagonalised leading to a hierarchy of mixings for different flavors and generations.~\cite{Panico:2015jxa}

Since the elementary fields are not in complete representations of G, the interactions between both sectors explicitly break the global symmetry of the SCFT, generating a potential for the NGBs at loop-level. The fermionic contributions to the potential can misalign the vacuum and trigger EWSB. As usual in composite Higgs models, we define:
\begin{equation}\label{eq-xi}
\xi\equiv \frac{v_{\rm SM}^2}{f^2} \ ,
\end{equation}
where $v_{\rm SM}=(\sqrt{2}G_F)^{-1/2}$. Since EW precision tests in general require $\xi\ll 1$, $\xi$ is a good expansion parameter.

\subsection{The global symmetry of the SCFT}
There are several conditions that guide us in the choice of G and H. 

First, the subgroup H must contain a custodial group SO(4)$\simeq$SU(2)$_L\times$SU(2)$_R$, to avoid large contributions to the $T$-parameter. Thus, since it must also contain the SM gauge symmetry,  
%H must contain a subgroup, that we call H$_{\rm min}$, with: 
$\textrm{H} \supset \textrm{H}_{\rm min}\equiv \rm{SU(3)_c}\times{\rm SU(2)_L}\times{\rm SU(2)_R}\times{\rm U(1)_X}$. The SM hypercharge is given by the linear combination $Y=T^3_R+\alpha X$, with $\alpha$ a real constant to be fixed later. Notice that, whereas the Higgs is taken as a bidoublet of SO(4), the SU(2)$_R$ charge of the other resonances, is not fixed. For example, only the linear combination corresponding to hypercharge is fixed for $S_3$: $Y=1/3$. 

Before going to the second set of conditions, we define our notation for the representations of the groups. In general we will denote them by their dimensions, whenever needed we will use a bar to distinguish a representation and its complex conjugate. We will use a calligraphic letter ${\cal R}$ for representations of G, a large letter ${\bf R}$ for representations of H, and a small ${\bf r}$ for representations of H$_{\rm min}$. Similarly we will use a small ${\bf n}$ for irreducible representations of SU(2).

As a second condition, the coset G/H must contain a set of generators transforming as ${\bf r}_H=({\bf 1},{\bf 2},{\bf 2})_0$ and ${\bf r}_{S_3}=(\bar {\bf 3},{\bf 3},{\bf n}_R)_{X}$ under H$_{\rm min}$, that will correspond to the Higgs and the leptoquark $S_3$, respectively. That is:
\begin{equation}
{\rm G/H}\sim {\bf R}_{\rm NGB} \sim {\bf r}_H \oplus {\bf r}_{S_3} \oplus \dots
\end{equation}
where the second relation stands for the decompositions of ${\bf R}_{\rm NGB}$ under H$_{\rm min}$. The dots are present because there could be other NGB states in G/H, besides $H$ and $S_3$

A suitable choice of groups G and H is given by
\begin{align}
& {\rm G}\simeq {\rm SO(13)} \ ,
\\
& {\rm H}\simeq {\rm SO(6)}\times{\rm SU(2)}^3\ .%\times{\rm SU(2)} \ ,
\end{align}
H can be obtained by following the pattern ${\rm SO(13)}\to {\rm SO(6)}\times{\rm SO(7)}$ and
${\rm SO(7)}\to{\rm SU(2)_1}\times{\rm SU(2)_2}\times{\rm SU(2)_R}$.

H$_{\rm min}$ is contained in H in the following way: first identifying ${\rm SU(2)_{L}}\equiv{\rm SU(2)_{1+2}}$, the diagonal subgroup contained in the product of the first and second SU(2), and also decomposing SO(6)$\simeq$SU(4)$\supset$SU(3)$_c\times$U(1)$_X$.

Once G and H are chosen, and the embedding of H$_{\rm min}$ in H is determined, it is straightforward to obtain the NGBs. In our case they decompose under H as:
\begin{align}
{\bf R}_{NGB}= {\bf R}_{S}\oplus{\bf R}_{R}\oplus{\bf R}_{H}=({\bf 6},{\bf 3},{\bf 1},{\bf 1})\oplus({\bf 6},{\bf 1},{\bf 2},{\bf 2})\oplus({\bf 1},{\bf 3},{\bf 2},{\bf 2}) \ . \label{eq-rPi}
\end{align}
The lowest dimensional irreducible representations of SO(6) decompose under SU(3)$\times$U(1) as:
\begin{align}
&{\bf 4}\sim {\bf 3}_{-1}\oplus{\bf 1}_3 \ , & {\bf 6}\sim {\bf 3}_{2}\oplus\bar {\bf 3}_{-2} \ , \nonumber \\
&{\bf 10}\sim {\bf 6}_2\oplus\bar {\bf 3}_{-2}\oplus{\bf 1}_{-6}\ , & {\bf 15}\sim {\bf 8}_0\oplus\bar {\bf 3}_{-4}\oplus{\bf 3}_4+{\bf 1}_0 \ .
\label{eq-so6-su3}
\end{align}
By using these results we obtain that under H$_{\rm min}$ the NGBs transform as
\begin{align}
& {\bf r}_{S}= (\bar {\bf 3},{\bf 3},{\bf 1})_{-2}+{\rm c.c.} \ ,\nonumber
\\
& {\bf r}_{R}= (\bar {\bf 3},{\bf 2},{\bf 2})_{-2}+{\rm c.c.} \ ,\nonumber
\\
& {\bf r}_{H}\oplus{\bf r}_{H_4}\sim ({\bf 1},{\bf 2},{\bf 2})_0\oplus({\bf 1},{\bf 4},{\bf 2})_0 \ ,
\label{eq-rNGB}
\end{align}
where c.c. stands for the complex conjugate representations, and where we have used that, for SU(2): ${\bf 2}\otimes{\bf 3}\sim{\bf 2}\oplus{\bf 4}$.~\footnote{We have used the subindex $R$ for the second line of Eq.~(\ref{eq-rNGB}) because, as we will show below, it leads to the leptoquarks usually denoted with the letter $R$, see for example the notation of Ref.~\cite{Dorsner:2016wpm}.} 

In order to obtain the proper hypercharge of $S_3$, we fix $\alpha=-1/6$, obtaining: 
\begin{equation}\label{eq-Y}
Y=T^{3R}-X/6 \ .
\end{equation}

Under G$_{\rm SM}$, ${\bf r}_{R}$ decomposes as the sum of two representations, corresponding to leptoquark doublets with different hypercharges. Therefore, besides the usual SM Higgs $H$, in the NGB spectrum there is a colorless fourplet $H_4$, and the leptoquarks $S_3$, $\tilde R_2$ and $\hat R_2$:
\begin{align}
&H\sim({\bf 1},{\bf 2})_{1/2} \ , \qquad H_4\sim({\bf 1},{\bf 4})_{1/2} \ ,\nonumber \\
&S_3\sim(\bar {\bf 3},{\bf 3})_{1/3} \ , \qquad \tilde R_2\sim({\bf 3},{\bf 2})_{1/6} \ ,\qquad \hat R_2\sim({\bf 3},{\bf 2})_{-5/6} \ .
\end{align}

Although SO(13) is the smallest simple group that we found, containing H$_{\rm min}$ and able to deliver $S_3$ and $H$ as NGBs, it also contains an extra pair of leptoquark doublets, as well as an extra colorless fourplet.

\subsubsection{Representations of the fermions}\label{sec-frep}
The SCFT operators ${\cal O}^{\rm SCFT}$ are in irreducible representations of G, whereas the elementary fermions only transform under G$_{\rm SM}$. The couplings of Eq.~(\ref{eq-pceff}) explicitly break G$_{\rm SM}\times$G to the diagonal subgroup. In order to understand several properties of this breaking, it is useful to add spurionic degrees of freedom in the elementary sector, embedding the SM fermions in the same irreducible representations of SO(13) as the operators of the SCFT mixing with them.~\footnote{These new elementary fermions, added to furnish complete representations of SO(13), are spurions, they do not correspond to propagating degrees of freedom.} There are several conditions that these representations must satisfy: to avoid an explicit breaking of G$_{\rm SM}$, they must contain components in the same representations under G$_{\rm SM}$ as the SM fermions, besides we require that they allow the usual Yukawa couplings with the Higgs, and finally we also require Yukawa interactions leading to $S_3 \bar q^c \ell$, with $q$ and $\ell$ the quark and lepton doublets. 

We find that the smallest representations of SO(13) in which the SM fermions can be embedded, are the following ones:
\begin{align}
&  {\cal R}_{q}= {\bf 286}\supset ({\bf 6},{\bf 3},{\bf 2},{\bf 2})={\bf R}_q \ , 
& {\cal R}_{u,d}= {\bf 286}\supset ({\bf 6},{\bf 1},{\bf 1},{\bf 3})={\bf R}_{u,d}  \ , \nonumber
\\
& {\cal R}_\ell= {\bf 78}\supset ({\bf 1},{\bf 3},{\bf 2},{\bf 2})={\bf R}_\ell \ , 
&\qquad {\cal R}_e= {\bf 78}\supset ({\bf 1},{\bf 1},{\bf 1},{\bf 3})={\bf R}_e \ ,
\label{eq-frep13-3}
\end{align}
where we have specified the component under H containing the SM fermions. We put all the generations in the same representations. 
We leave a description of the lowest dimensional representations of SO(13), and their decompositions under H and H$_{\rm min}$, for App.~\ref{ap-SO(13)}. 

The following embedding also contains a state with the same quantum numbers as $q$:
%\begin{equation}
${\cal R}_{q}= {\bf 286}\supset {\bf R}_{q'}=({\bf 15},{\bf 1},{\bf 2},{\bf 2}) \ $.
%\end{equation}
However, unless $u$ and $d$ are embedded in higher dimensional representations of SO(13), ${\bf R}_{q'}$ does not generate the usual Yukawa couplings. Besides, it induces ${\rm LQ}qq$ interactions that, as we will discuss in sec.~\ref{sec-sqq}, can induce proton decay.~\footnote{We will use $LQ$ generically for leptoquarks.} For this reason we will assume that the mixing with the component ({\bf 15},{\bf 1},{\bf 2},{\bf 2}) is very small, and we will not consider it in our analysis (in the appendices we will show its effect on the potential). 

There are smaller irreducible representations to embed the SM fermions, but they not satisfy all the conditions discussed in the beggining of this section. It is also possible to embed $\ell$ in ${\bf 286}\supset ({\bf 1},{\bf 3},{\bf 2},{\bf 2})$, but for simplicity we will work only with the embeddings of Eq.~(\ref{eq-frep13-3}). 

%The quark doublet $q_L$ is embedded in two different representations, $q^u$ ($q^d$) generates the mass of the up- (down-) type quarks.

We will call $\Psi_f$, with $f=q,u,d,\ell,e$, to the chiral fermion obtained after the embedding of the elementary fermion $f$ into a representation of SO(13). For example: $\Psi_q$ will be an elementary Left-handed fermion in the representation {\bf 286}, where only the components corresponding to the SM quark doublet, {\it i.e.}: in the representation $({\bf 3},{\bf 2})_{1/6}$ of the SM group, are dynamical, and the other components are not dynamical.

\subsubsection{A symmetry to forbid baryon decay}\label{sec-sqq}
The interactions involving two quarks and one leptoquark can induce baryon decay. In our model there are ${\rm LQ}qq$ interactions at the TeV scale that make the theory phenomenologically unacceptable. However these interactions can be forbidden by imposing a ${\it Z}_2$-symmetry from SO(13), as: $P=e^{iT_P\pi/2}$, with $T_P$ a generator of the SO(6) subgroup, see App.~\ref{ap-SO(13)}. In the representation {\bf 13} of SO(13), choosing a suitable basis, $P$ can be written as a block diagonal matrix: $P={\rm diag}(I_7,-I_6)$, where $I_6$ and $I_7$ are the identity in SO(6) and SO(7), respectively.~\footnote{See \cite{Gripaios:2014tna} for a similar symmetry in a factorizable group.} As an example, fields in the fundamental representation of SO(6) are odd under $P$, as the quarks in ${\bf R}_{q,u,d}$ and the leptoquarks, whereas fields in the singlet or adjoint representation of SO(6) are even, as the leptons and the quarks in ${\bf R}_{q'}$. This symmetry forbids the interactions ${\rm LQ}qq$ and ${\rm LQ}q'q'$, however it allows interactions ${\rm LQ}q'q$, with $q$ a quark in ${\bf R}_{q,u,d}$ and $q'$ in ${\bf R}_{q'}$. To forbid transitions mediated by the last operator, the projection on ${\bf R}_{q'}$ must be suppressed, thus we take $\lambda_{q'}=0$ and neglect its effect in the following.

\subsection{Flavor}
As is well known, to explain the SM fermion masses and mixing in anarchic partial compositeness, the mixing $\epsilon$ of Eq.~(\ref{eq-epsilon}) must satisfy some relations. In cases where each elementary multiplet mixes with just one composite resonance, these conditions are~\cite{Agashe:2004cp,Csaki:2008zd,Agashe:2008uz}:~\footnote{If an elementary fermion interacts with several operators of the SCFT, there can be more freedom~\cite{Csaki:2008zd,Panico:2015jxa}. If at low energies one of these couplings is much larger than the others, for example because they have different scaling dimensions, considering just the leading one gives a good approximation.}
\begin{align}
&\epsilon_{q1}\sim\lambda_C^3 \epsilon_{q3}\ , 
\qquad 
\epsilon_{u1}\sim \frac{y_u^{\rm SM}}{\lambda_C^3g_*\epsilon_{q3}} \ , 
\qquad
\epsilon_{u2}\sim \frac{y_c^{\rm SM}}{\lambda_C^2g_*\epsilon_{q3}}  \ , 
\qquad
\epsilon_{u3}\sim \frac{y_t^{\rm SM}}{g_*\epsilon_{q3}} \ ,
\nonumber \\
&\epsilon_{q2}\sim\lambda_C^2 \epsilon_{q3}\ ,
\qquad
\epsilon_{d1}\sim \frac{y_d^{\rm SM}}{\lambda_C^3g_*\epsilon_{q3}} \ ,
\qquad
\epsilon_{d2}\sim \frac{y_s^{\rm SM}}{\lambda_C^2g_*\epsilon_{q3}} \ ,
\qquad
\epsilon_{d3}\sim \frac{y_b^{\rm SM}}{g_*\epsilon_{q3}} \ ,
\label{eq-mixq}
\end{align}
where $\lambda_C$ is the Cabibbo angle and $y_f^{\rm SM}=m_f/v_{SM}$. The first column leads to the CKM matrix, the other columns lead to the quark masses. The only free parameters are $g_*$ and $\epsilon_{q3}$.

The linear couplings of the leptons can not be fixed as in the quark sector, a mechanism generating neutrino masses must be chosen first. Flavor constraints can be minimised if Left- and Right-handed couplings of charged fermions are taken of the same order:~\cite{Panico:2015jxa}
\begin{equation}
\epsilon_{\ell 1}\sim\epsilon_{e1}\sim \sqrt{y_{e}^{\rm SM}/g_*} \ ,\qquad
\epsilon_{\ell 2}\sim\epsilon_{e2}\sim \sqrt{y_{\mu}^{\rm SM}/g_*} \ ,\qquad
\epsilon_{\ell 3}\sim\epsilon_{e3}\sim \sqrt{y_{\tau}^{\rm SM}/g_*} \ ,
\label{eq-mixl}
\end{equation}
we will consider this choice in the present work. In this case the unitary matrices diagonalising the charged lepton mass matrix have hierarchical angles, thus the angles of the PMNS matrix are generated in the neutrino sector, see Refs.~\cite{Agashe:2008fe,Panico:2015jxa,DaRold:2017xdm} for these scenarios.

\subsection{Constraints}\label{sec-constraints}
We consider first the most important effects on the oblique parameters and $Zb\bar b$ couplings related with composite grand unification. As we will discuss, due to the presence of an extended scalar sector, there are new contributions to the $\hat T$-parameter, that are absent in the MCHM containing a single scalar. However we will show that these contributions are suppressed for small $\xi$.~\cite{Babu:2009aq}  

A is well known, the Higgs potential of the SM has an accidental SO(4) global symmetry, spontaneously broken to the custodial  symmetry SO(3)$_c$ by the Higgs vacuum expectation value (vev), that is behind the relation $\rho\simeq 1$. We summarize first how this SO(3)$_c$ is preserved in the MCHM~\cite{Agashe:2004rs,Mrazek:2011iu}, and after that we discuss it in our model. In the MCHM the Higgs is the NGB in the coset SO(5)/SO(4), with the strongly interacting sector having an exact SO(4)-symmetry. The SM gauges an SO(4)$_g$ subgroup of SO(5), besides this gauging induces a potential for the NGB, eventually misaligning the vacuum. The misalignment happens if the group preserved by the vacuum, SO(4)$_{\rm vac}$, is different from SO(4)$_g$. However, two different SO(4)'s embedded in SO(5) always share a common SO(3) subgroup. In the present model the color singlets $H$ and $H_4$ are in the coset SO(7)/[SO(4)$\times$SO(3)], they transform as a $({\bf 4},{\bf 3})$ of the invariant subgroup. The SM gauges (a subgroup of) an [SO(4)$\times$SO(3)]$_g$ subgroup of SO(7), generating a potential for the NGB and eventually misaligning the vacuum. The misalignment happens if the group preserved by the vacuum, [SO(4)$\times$SO(3)]$_{\rm vac}$, is different from [SO(4)$\times$SO(3)]$_g$. We find three possibilities for the misalignment, that depend on which subgroup is shared by these two groups: (a) an SO(3) subgroup, in this case only $H$ has a vev, (b) an SO(2) subgroup, in this case $\langle H_4\rangle$ is annihilated by the same generator as $\langle H\rangle$, and both vevs have the same charge under $T^3_L$, and (c) the trivial subgroup, as happens for generic vevs $\langle H\rangle$ and $\langle H_4\rangle$ that do not satisfy the conditions of case (b). Case (a) is the most favorable one, containing a custodial symmetry, whereas case (c) is not compatible with the phenomenology, since there is no massless photon in the spectrum. In sec.~\ref{subsec-ewsb} we will show that, in our model, case (b) is realised, since the presence of the Higgs vev triggers a vev of the neutral component of $H_4$: $v_4$~\cite{Babu:2009aq}. Case (c) is also possible in our model. A non-vanishing $v_4$ modifies the $\rho$-parameter as: $\rho\simeq 1-6v_4^2/v^2$. Ref.~\cite{Babu:2009aq} has shown that the constraints on $\rho$ require, at 3 $\sigma$ level, $v_4\lesssim 2.5$~GeV.~\footnote{It has also shown that there are positive contributions to $\rho$ induced by the splitting of the $H_4$ components.} We will show that $v_4$ is suppressed compared with $v$ by: $v_4\sim \xi v/2$, leading to $\rho\sim 1+{\cal O}(\xi^2)$. By considering just this contribution to $\rho$, and neglecting corrections to other EW parameters, $\xi\lesssim 0.02-0.04$, increasing the amount of tuning compared with the usual MCHM, that requires $\xi\lesssim 0.1-0.3$.

As we will show, the vev of $H_4$ is generated by a term in the potential of the form $(H^2H^\dagger H_4^\dagger)$. We have searched for symmetries that could prohibit this term, relaxing the bounds from $\rho$. An example would be a parity transformation such that $H$ and $H_4$ have different eigenvalues under this operation, for example $\pm 1$. We have found that there is no non-trivial element in the algebra of SO(7) having $H$ and $H_4$ as eigenvectors. Since the exponential map is surjective for SO(7), this result covers all the possibilities. Thus there are no symmetries inside SO(7) that could forbid the cited term in the potential. Extending the group to O(7) does not offer new solutions.

Corrections from new physics to $Zb_L\bar b_L$ coupling can not be larger than $\sim 0.25\%$. In composite Higgs models with partial compositeness, in the simple framework of one scale and one coupling in the sector of resonances, the tree-level corrections can be estimated as $\delta g_{b_L}/g\sim \xi \epsilon_{q3}^2$.
For $f\sim800$~GeV, $\delta g_{b_L}$ is usually too large. However it is possible to protect the $Z$-couplings with a discrete subgroup of the custodial symmetry, a parity $P_{LR}$, ensuring that $\delta g_{b_L}$ is sufficiently suppressed.~\cite{Agashe:2006at} This symmetry requires embedding $q_L$ in a $({\bf 2},{\bf 2})_{2/3}$ of SU(2)$_L\times$SU(2)$_R\times$U(1)$_X$. From Eqs.~(\ref{eq-Y}) and (\ref{eq-frep13-3}), one can see that, by choosing ${\cal R}_q={\bf 286}$, $q_L$ is embedded in a $({\bf 2},{\bf 2})_{-1/3}$ of SU(2)$_L\times$SU(2)$_R\times$U(1)$_X$. Thus extra tuning could be needed, with this choice of ${\cal R}_q$, to pass the constraints from $Zb\bar b$.

In secs. \ref{sec-potential} and~\ref{sec-pheno} we will show the prediction of $v_4$ and $\delta g_{b_L}$ in our model, as well as the tuning.

One possibility to avoid too large $\delta g_{b_L}$ is to find an ${\cal R}_q$ containing a $({\bf 2},{\bf 2})_{2/3}$ for the Left-handed quarks. The smallest SO(13) representation that we have found with this property is: ${\cal R}_q=\overline{\bf 715}$, that contains an ${\bf R}_q=({\bf 15},{\bf 3},{\bf 2},{\bf 2})$, allowing the proper embedding of $q_L$. In this case one can choose, for example, ${\cal R}_u={\bf 78}$, leading to the right Yukawa coupling with the Higgs. Given the large dimension of ${\cal R}_q$, we have not pursued this analysis.

Other strong constraints in this kind of theories arise from neutron-antineutron oscillations. This process is induced by operators of dimension 9, involving six quarks of the first generation.~\cite{Phillips:2014fgb} However it has been shown that in the framework of anarchic partial compositeness, with a compositeness scale in the range of few TeV, the Wilson coefficients of these operators are sufficiently suppressed.~\cite{Gripaios:2014tna}

In sec.~\ref{sec-pheno} we will comment on other phenomenological constraints, as direct searches at collider, flavor transitions and lepton flavor universality violation.

\section{Effective theory}\label{eff-th}
At energies below $m_*$ the heavy resonances of the SCFT can be integrated-out, leading to an effective theory with the SM degrees of freedom, plus the NGBs. Given the symmetries and the fermionic representations, many properties of this effective low energy description are fixed. 

By using the CCWZ formalism~\cite{Coleman:1969sm,Callan:1969sn}, one can build an effective Lagrangian that, although superficially looks only H-invariant, is G-invariant after embedding the SM fermions in representations of G. One of the main objects for this construction is the NGB matrix, that is defined as:
\begin{equation}
U=e^{i\sqrt{2}\Pi/f}\ ,\qquad \Pi=\Pi^{\hat a}T^{\hat a}
\end{equation}
with $T^{\hat a}$ the broken generators of G, $\Pi^{\hat a}$ the associated NGB fields and $f$ their decay constant. In fact, since $\{T^{\hat a}\}$ spans a reducible representation of H, see Eq.~(\ref{eq-rPi}), there are three independent decays constants: $f_S$, $f_R$ and $f_{H}$.
%, that are taken to be of the same order, and are collectively denoted as $f$ 

$U$ transforms under a transformation ${\cal G}\in$G as: $U\to{\cal G}U{\cal H}^{-1}$, where ${\cal H}$ is an element of H that depends on ${\cal G}$ and $\Pi$. As usual in the CCWZ formalism, the kinetic term of the NGBs can be written as
\begin{equation}\label{eq-L2NGBs}
{\cal L}\supset \sum_{{\bf R}_{\rm NGB}}\frac{f_{\bf R}^2}{4}d_\mu^{\hat a_{\bf R}}d^{\mu\hat a_{\bf R}} \ ,
%{\cal L}\supset \frac{f^2}{4}d_\mu^{\hat a}d^{\mu\hat a} \ ,
\end{equation}
where $d_\mu$ can be defined from the Cartan-Maurer form: $iU^\dagger D_\mu U=d_\mu^{\hat a}T^{\hat a}+e_\mu^aT^a$, and $D_\mu$ is the usual covariant derivative.  
%Since the NGBs transform with a reducible representation of H, Eq.~(\ref{eq-rPi}), one has to include a sum over ${\bf R}$. 
For simplicity we will take the same numerical value for all the decay constants, calling it $f$.

If we assume that the vacuum preserves an electromagnetic U(1) symmetry, evaluating the NGBs $H$ and $H_4$ in their vevs, $v$ and $v_4$, Eq.~(\ref{eq-L2NGBs}) generates a mass term for the $Z$ and $W$s, with the matching:
\begin{equation}\label{eq-vSM}
v_{\rm SM}^2=(246{\rm GeV})^2=\frac{f^2}{6}\left[
9\sin^2\left(\frac{3v_4}{\sqrt{2}f}\right)
+2\sin^2\left(\frac{v}{f}+\frac{v_4}{\sqrt{2}f}\right)
+\sin^2\left(\frac{2v}{f}-\frac{v_4}{\sqrt{2}f}\right)
\right] \ .
\end{equation}

Eq.~(\ref{eq-vSM}) is invariant under the following combined transformation: $v\to-v$ and $v_4\to-v_4$. Less obvious, but straightforward to check, it is also invariant under the combined transformation: $v\to v+2\pi f/3$ and $v_4\to v_4+\sqrt{2}\pi f/3$. Besides it has period $\pi f$ and $\sqrt{2}\pi f$ in the variables $v$ and $v_4$.
We show $v_{\rm SM}/f$ as function of $v/f$ and $v_4/f$ in the right-panel of Fig.~\ref{fig-potential}.

\subsection{Sector of fermions}\label{sec-fermions}
We describe now the interactions between the NGBs and the fermion fields. To write an invariant Lagrangian including fermions, we will employ the CCWZ formalism, here we summarise the basic tools that will be needed for our model.
By dressing with $U^\dagger$ a field $\Psi$ that is in an irreducible representation ${\cal R}$ of G, and projecting it on any irreducible representation ${\bf R}$ of H contained in ${\cal R}$, it is possible to obtain fields that under the action of G transform in reducible representations of H:
\begin{equation}
\tilde \Psi^{\bf R}=P_{\bf R}(U^\dagger\Psi) \ ,
\label{eq-upsi}
\end{equation}
where $P_{\bf R}$ is the projector from ${\cal R}$ to ${\bf R}$.

A product of dressed fields decomposes under transformations of H as the sum of irreducible representations of H, according to ${\bf R}_1\otimes{\bf R}_2\sim \oplus_j{\bf R}_j$. To shorten notation we will denote $({\bf R}_1\otimes{\bf R}_2)_{{\bf R}_j}\equiv P_{{\bf R}_j}({\bf R}_1\otimes{\bf R}_2)$, where the subindex ${\bf R}_j$ in the l.h.s. indicates that the product is projected onto the representation ${\bf R}_j$ of H. In particular, we will be interested in the presence of H-singlets: ${\bf R}_j={\bf 1}$, that will correspond to G-invariant terms. In the present case, the product $({\bf R}_1\otimes{\bf R}_2)$ will contain singlets only if ${\bf R}_1\sim{\bf R}_2$, therefore it is enough to consider this case for invariants arising from the product of two representations.

Following the usual algorithm described below Eq.~(\ref{eq-upsi}), it is possible to write G-invariant terms containing the usual Yukawa and leptoquark interactions, as well as an infinite series of terms with higher powers of the NGBs. For quark bilinears, since all the quarks have been embedded in the representation {\bf 286}, one has to sum over all the irreducible representations of H contained in {\bf 286}: $\sum_{{\bf R}\subset {\bf 286}}\left[\bar{\tilde \Psi}_q^{\bf R} (c^u_{\bf R}\tilde\Psi_u^{\bf R}+c^d_{\bf R}\tilde\Psi_d^{\bf R})\right]_{\bf 1}$. In this expression $c^u_{\bf R}$ and $c^d_{\bf R}$ are coefficients independent of the fields. Expanding to first order in the NGBs and putting to zero the non-dynamical fermions, it is straightforward to obtain the usual Yukawa interactions of the up- and down-type quarks. The same results apply for the leptons, now embedded in the representation {\bf 78} of G: $\sum_{{\bf R}\subset {\bf 78}}c^e_{\bf R}(\bar{\tilde \Psi}_\ell^{\bf R}\tilde \Psi_e^{\bf R})_{\bf 1}$. One can also write invariants with quarks and leptons, that will lead to leptoquark interactions. The common H-representations in the decomposition of {\bf 286} and {\bf 78} can be read in Eq.~(\ref{eq-frep13-1}) of App.~\ref{ap-SO(13)}: ${\bf R}\sim({\bf 1},{\bf 3},{\bf 2},{\bf 2}),({\bf 6},{\bf 3},{\bf 1},{\bf 1})$, thus leptoquark interactions can be obtained from invariants as: $\sum_{\bf R}c^{q\ell}_{\bf R}(\bar{\tilde \Psi}_{q^C}^{\bf R}\tilde \Psi_\ell^{\bf R})_{\bf 1}$, where the sum is over the common ${\bf R}$'s. It is straightforward to check that, to first order in the NGBs, only the usual Yukawa interactions with the Higgs, as well as interactions with $S_3$, are generated, no more interactions are present to this order. In sec.~\ref{sec-pheno} we will show explicitly the leading terms in an expansion in powers of the NGBs.

By dressing the fermions with U, the effective Lagrangian quadratic in the fermions can be written as
%\begin{equation}\label{eq-Leff1}
%{\cal L}_{\rm eff}=\sum_{f=q^u,q^d,u,d,\ell,e}\bar\Psi_f\ Z_f \pslash\ \Psi_f+\sum_{f,f'=q^u,q^d,u,d,\ell,e}\sum_{\bf R}[\bar{\tilde\Psi}_f^{\bf R}\ \Pi_{f,f'}^{\bf R}(p)\ \tilde\Psi_{f'}^{\bf R}]_{\bf 1} \ , 
%\end{equation}
\begin{align}\label{eq-Leff1}
&{\cal L}_{\rm eff}=\sum_f\bar\Psi_f\ Z_f \pslash\ \Psi_f+\sum_{f,f'}\sum_{\bf R}[\bar{\tilde\Psi}_f^{\bf R}\ \Pi_{f,f'}^{\bf R}(p)\ \tilde\Psi_{f'}^{\bf R}+\bar{\tilde\Psi}_f^{\bf R}\ \Pi_{f,f'^C}^{\bf R}(p)\ \tilde\Psi_{f'^C}^{\bf R}]_{\bf 1} +{\rm h.c.}\ , \nonumber \\
&f,f'=q,u,d,\ell,e \ , 
\end{align}
where a sum over generations is understood.~\footnote{We have not included the neutrino sector, since it depends on the nature of the neutrino, and is thus more model dependent, but it is straightforward to include it.} The first term is the elementary kinetic term, whereas the second and third terms are the contributions from the SCFT. The third sum contains an elementary fermion $f$ and a charge conjugate elementary fermion: $f'^{C}$. $\Pi_{f,f'}^{\bf R}(p)$ and $\Pi_{f,f'^C}^{\bf R}(p)$ are form-factors depending on momentum that contain the information arising from the integration of the massive resonances, although they contain an index ${\bf R}$, they are numbers under transformations of H, they do not depend on the NGBs. Since the dominant interactions between the elementary fermions and the SFCT are linear, the $\epsilon_f$ dependence of the form factors $\Pi_{f,f'}^{\bf R}(p)$ can be factorised as: $\Pi_{f,f'}^{\bf R}(p)=\epsilon_f\epsilon_{f'}\hat\Pi_{f,f'}^{\bf R}(p)$, with $\hat \Pi_{f,f'}^{\bf R}(p)$ depending on momentum and on the flavor anarchic parameters of the SCFT only. When three generations are considered, $\epsilon_{f}$ and $\hat \Pi_{f,f'}^{\bf R}(p)$ become matrices in flavor space.~\footnote{By a field redefinition the matrices $\epsilon_{f}$ can be taken diagonal.~\cite{Panico:2015jxa}} If $f$ and $f'$ ($f'^C$) have the same chirality, a $\pslash$ can be factorised from $\Pi_{f,f'}^{\bf R}(p)$ ($\Pi_{f,f'^C}^{\bf R}(p)$), in the following we will assume this factorization has been done. In App.~\ref{ap-2site} we give explicit expressions of the form factors in the case of a two-site theory.

Evaluating the NGBs on its vev, and keeping only the dynamical elementary fermions, ${\cal L}_{\rm eff}$ reduces to
\begin{equation}\label{eq-Leff2}
{\cal L}_{\rm eff}=\sum_{f=u,d,e}\left[\bar f_L M_f f_R +{\rm h.c.}+\sum_{X=L,R}\bar f_X \pslash(Z_f+\Pi_{f_X})f_X\right]
\end{equation}
%where generation indices are understood, and 
where we have assumed that only $H$ and $H_4$ have vevs.

The correlators $\Pi_f$ and $M_f$ can be obtained from Eq.~(\ref{eq-Leff1}) as:
\begin{align}
&\Pi_{u_L}=\sum_{\bf R} i_{q,u_L}^{\bf R}\Pi_{q,q}^{\bf R} \ ,%+ i_{q',u_L}^{\bf R}\Pi_{q'}^{\bf R})\ , 
& \Pi_{u_R}= \sum_{\bf R} i_{u,u_R}^{\bf R}\Pi_{u,u}^{\bf R}\ , \nonumber \\
&\Pi_{d_L}=\sum_{\bf R} i_{q,d_L}^{\bf R}\Pi_{q,q}^{\bf R} \ ,%+ i_{q',d_L}^{\bf R}\Pi_{q'}^{\bf R})\ , 
& \Pi_{d_R}= \sum_{\bf R} i_{d,d_R}^{\bf R}\Pi_{d,d}^{\bf R}\ , \nonumber \\
&\Pi_{e_L}=\sum_{\bf R} i_{\ell,e_L}^{\bf R}\Pi_{\ell,\ell}^{\bf R} , 
& \Pi_{e_R}= \sum_{\bf R} i_{e,e_R}^{\bf R}\Pi_{e,e}^{\bf R}\ ,
\nonumber \\
&M_{f}=\sum_{\bf R} j_{f}^{\bf R}\Pi_{q,f}^{\bf R} \ , 
\quad f=u,\ d,\ ,
&M_{e}=\sum_{\bf R} j_{e}^{\bf R}\Pi_{\ell,e}^{\bf R} \ ,
\label{eq-correlators}
\end{align}
the functions $i_{f,f'}^{\bf R}$ and $j_{f}^{\bf R}$ can be computed to all orders in $v/f$ and $v_4/f$. Defining $s_z=\sin z$, $c_z=\cos z$, $x=v/f+v_4/(f\sqrt{2})$ and $y=v/f-\sqrt{2}v_4/f$, we show our results for the quarks in table \ref{t-q1}.% and \ref{t-q2}, whereas for the leptons they are in table \ref{t-l}.

\begin{table}[ht]
  \begin{tabular}{|c|c|c|c|c|}
    \hline\rule{0mm}{5mm}
$i$  & ({\bf 6},{\bf 3},{\bf 1},{\bf 1})  & ({\bf 6},{\bf 1},{\bf 1},{\bf 3}) & ({\bf 6},{\bf 1},{\bf 3},{\bf 1}) & ({\bf 6},{\bf 3},{\bf 2},{\bf 2}) \\[5pt]
\hline \rule{0mm}{5mm}
 $i_{q,u_L} $  &$ \frac13 s_{x+y}^2$ &$ \frac16 (s_x - s_{x + y})^2$ & $\frac16 (s_x + s_{x + y})^2$ & $\frac16 (3 + c_{2x} + 2 c_{2x +2 y})$ \\
    \hline\rule{0mm}{5mm}
$i_{q,d_L}$    &  $ \frac16 s_{2x}^2$ & $\frac{1}{12} (s_{2x} - s_y)^2 $ & $
\frac{1}{12}(s_{2x} + s_y)^2$ & $  \frac{1}{12}(9 + 2c_{4x} + c_{2y})$\\
    \hline\rule{0mm}{5mm}
 $i_{u_R} $  &   $\frac12 s_x^2 s_y^2$ & $c_x^2 s_{y/2}^4 $&$ c_x^2 c_{y/2}^4$ & $\frac12 (1
- c_{2x} c_{y}^2)$ \\
    \hline\rule{0mm}{5mm}
 $i_{d_R}$   & $ \frac12 s_{x}^4$ & $\frac{1}{16} (c_{2x}-2 c_{y} +1)^2$ & $\frac{1}{16} (c_{2x}+2 c_y+1)^2$ & $\frac18 (-c_{4x}-2 c_{2 y}+3)$  \\
    \hline\rule{0mm}{5mm}
 $j_u$  &  $\frac{1}{\sqrt{6}} (s_x s_y s_{x+y})$ & $ \sqrt{\frac23} c_x s_{y/2}^3
c_{x+y/2}$ & $-\sqrt{\frac23} c_x c_{y/2}^3 s_{x+y/2}$ &$
\frac{1}{\sqrt{6}} (c_y s_{2x+y})$ \\
\hline\rule{0mm}{5mm}
$j_d$ & $\frac{1}{\sqrt{3}} (c_x s_x^3)$ & $\frac{1}{4 \sqrt{3}} (c_y - c_x^2)(s_{2x} - s_y)$ & $-\frac{1}{4\sqrt{3}} (c_y + c_x^2)(s_{2x}+s_y)$ & $\frac{1}{4 \sqrt{3}}(s_{4x} + s_{2y})$ \\ 
    \hline 
  \end{tabular}
\caption{Invariants evaluating the NGBs in their vevs, with $s_z=\sin z$, $c_z=\cos z$, $x = v/f + v_4/(f\sqrt{2})$ and $y = v/f - \sqrt{2} v_4/f$. The columns are associated to representations ${\bf R}$ present in the decomposition of ${\bf 286}$ under H.}
\label{t-q1}
\end{table}

The fermionic spectrum can be obtained by computing the equations of motion from the Lagrangian (\ref{eq-Leff2}), it is given by:
\begin{equation}
{\rm zeroes}\left[p^2(Z_{f_L}+\Pi_{f_L})(Z_{f_R}+\Pi_{f_R})-|M_f|^2\right] \ .
\end{equation}

\section{Potential}\label{sec-potential}
The SM fields explicitly break the global symmetry of the SCFT. Keeping only the dynamical SM fields, and putting to zero the spurions that were introduced to obtain full representations of SO(13), a potential for the NGBs is generated.

The fermion contribution to the Coleman-Weinberg potential at one loop can be written as:
\begin{equation}\label{eq-VCW}
V= -\int \frac{d^4p}{(2\pi)^4}\log{\rm det}\ {\cal K}\ ,
\end{equation}
where ${\cal K}$ is the ``matrix'' in the Lagrangian of Eq.~(\ref{eq-Leff2}), when it is written as: ${\cal L}_{\rm eff}=\bar F{\cal K}F$, with $F^t=(f,f^C)$ and $f$ the chiral fermions of the SM. ${\cal K}$ has the SU(3)$_c$ and SU(2)$_L$ indices of the fermions in $f$ and it depends on the NGBs. Since in the anarchic approach $q$ and $u$ of the third generation have the largest interactions with the SCFT, they dominate the contributions to the potential, thus we will not consider the effect of the other fermions for the calculation of $V$. In this case ${\cal K}$ is a matrix of dimension nine. To shorten notation, in this section we will simply use $q$ and $u$ for the quarks of the third generation, without writing the generation index.

For simplicity we will not consider the contribution of the gauge fields to the potential, although it is straightforward to include it. Since the interactions of the third generation of fermions are usually stronger than the gauge ones, we expect the gauge fields to give a subdominant correction to the potential.~\footnote{If $g_{\rm SM}/g_*\sim \epsilon_{q3,u3}$, the gauge contributions to the potential are expected to be of the same size as the contribution of the fermions, thus they must be included.}

We have not been able to resum the matrix $U$ when all the NGBs are present. One can perform an expansion of $V$ in powers of the NGBs. In App.~\ref{ap-potential} we describe a method for this perturbative expansion. To fourth order in $\Pi$ the potential can be written as:
\begin{equation}\label{eq-V4}
V\simeq V_2+V_3+V_4+{\cal O}(\Pi^5)
\end{equation}
where $V_n$ is of order $n$ in the NGB,
\begin{equation}\label{eq-V42}
V_2 = m_{S_3}^2 |S_3|^2 + m_{\tilde R_2}^2 |\tilde R|^2 + m_{\hat R_2}^2 |\hat R|^2 + m_{H}^2 |H|^2 + m_{H4}^2 |H_4|^2 \ ,
\end{equation}
\begin{equation}\label{eq-V43}
V_3 = m_1 S_3 \tilde R H^\dagger + m_2 S_3 \hat R H + m_3 S_3 \tilde R H_4^\dagger + m_4 S_3 \hat R H_4 +\ {\rm h.c.}  \ ,
\end{equation}
and
\begin{equation}
V_4 =  V_4^H + V_4^{\rm LQ} + V_4^{H{\rm LQ}} =\sum_{j=1,\dots 49} c_j\ (\Phi_{a_j}\Phi_{b_j}\Phi_{d_j}\Phi_{e_j})\ ,
\label{eq-V44}
\end{equation}
where the superindex in $V_4$ specifies the kind of NGBs, $H$ for color singlets and ${\rm LQ}$ for leptoquarks, $c_j$ is a quartic coupling and $(\Phi_{a_j}\Phi_{b_j}\Phi_{d_j}\Phi_{e_j})$ is a SM singlet of fourth order in the NGBs. Since there are forty-nine quartic terms, we list them in App.~\ref{ap-quartic}. There are eight invariants in $V_4^H$, one involving only $H$, two with $H_4$ and five with $H$ and $H_4$, twenty-one in $V_4^{\rm LQ}$ and twenty in $V_4^{H{\rm LQ}}$, involving two fields that are color singlets and two leptoquarks. For details see App.~\ref{ap-quartic}. The coefficients of Eqs.~(\ref{eq-V42})-(\ref{eq-V44}) can be expressed as momentum integrals of the form factors of the effective theory. We show explicit expressions for the quadratic and cubic couplings in App.~\ref{ap-match}, the quartic ones involve very long expressions, therefore we only show some of them in the limit of large $Z_f$.

For the analysis of EWSB of the next section, it will be useful to know explicitly $V_4^H$:
\begin{align}
V_4^H=& c_{1} (H^\dagger H)_1^2+c_{2}(H_4^2)_3(H_4^{\dagger 2})_3+c_{3}(H_4^2)_7(H_4^{\dagger 2})_7+c_{4}(H^\dagger H)_1(H_4^{\dagger}H_4)_1
\nonumber\\&
+c_{5}(H^\dagger H)_3(H_4^{\dagger}H_4)_3+c_{6}H^{\dagger 2} H H_4+c_{7}H_4^{\dagger 2} H_4 H+c_{8}(H^2)_3(H_4^{\dagger 2})_3 +{\rm h.c.},
\label{eq-V4H}
\end{align}
where the h.c. is required for the last three terms. The subindex in the parenthesis shows the dimension of the SU(2)$_L$ representation chosen from the product of fields, as explained below Eq.~(\ref{eq-upsi}). Other quartic invariants depending on these fields can be written in terms of the ones shown in Eq.~(\ref{eq-V4H}).

It is also useful to study the potential expanding it in powers of the degree of compositeness of the fermions: $\epsilon_f$. To ${\cal O}(\epsilon_f^4)$, it can be written as~\cite{Panico:2012uw}
\begin{align}\label{eq-Veps}
V&\simeq \frac{m_*^4}{16\pi^2}\left[\epsilon_q^2 F_q^{(2)}(\Pi/f)+\epsilon_u^2 F_u^{(2)}(\Pi/f)
%\right. \nonumber\\&\left.
+\epsilon_q^4 F_q^{(4)}(\Pi/f)+\epsilon_u^4 F_u^{(4)}(\Pi/f)+\epsilon_q^2\epsilon_u^2 F_{qu}^{(4)}(\Pi/f)\right]
\end{align}
where $F_f^{(n)}$ are functions of the NGBs arising from the invariants, thus depending on the representations of the fermions. 

By using the expansion of Eq.~(\ref{eq-Veps}) one can estimate the size of the coefficients of Eqs.~(\ref{eq-V42})-(\ref{eq-V44}). Up to accidental cancellations of leading terms, we obtain: 
%for the quadratic, cubic and quartic coefficients of Eq.~(\ref{eq-V4}):
\begin{equation}\label{eq-eV4}
m_\Phi^2\sim \epsilon_f^2 \frac{m_*^4}{16\pi^2f^2} \ ,\qquad
m_n\sim \epsilon_f^2 \frac{m_*^4}{16\pi^2f^3} \ ,\qquad
c_j\sim \epsilon_f^2 \frac{m_*^4}{16\pi^2f^4} \ .
\end{equation}

\subsection{EWSB}\label{subsec-ewsb}
Successful EWSB requires a non-trivial minimum, where a U(1) symmetry associated with electromagnetism is preserved. Relying on the fourth order expansion of the potential, we demand:
\begin{equation}\label{eq-min}
m_H^2<0\ , \qquad m_\Phi^2>0 \ ,\qquad \Phi=S_3,\ \tilde R_2,\ \hat R_2,\ H_4 \ ,
\end{equation}
as well as positive quartic couplings stabilising the minimum. The presence of the coupling $c_{6}$ induces a vev of the neutral component of $H_4$~\cite{Babu:2009aq}:
\begin{equation}
v^2\simeq -\frac{m_H^2}{c_{1}} \ ,\qquad v_4\simeq -c_{6}v\left[\frac{v^2}{2m_4^2}+{\cal O}\left(\frac{c_iv^4}{m_4^4}\right)\right]
\label{eq-vmin}
\end{equation}

As usual in composite Higgs models, in the absence of tuning: $v\sim f$. However, as discussed in sec.~\ref{sec-constraints}, EW precision observables require $\xi\ll 1$, in this case $v_4$ is suppressed by a factor $\xi$ compared with $v$. 

Making use of table~\ref{t-q1}, it is possible to obtain the one-loop potential of Eq.~(\ref{eq-VCW}) to all orders in $v$ and $v_4$:
\begin{equation}
V=-N_c\int\frac{d^4p}{(2\pi)^4}\left\{\log\left[p^2(Z_{u_L}+\Pi_{u_L})(Z_{u_R}+\Pi_{u_R})-|M_u|^2\right]+\log \left[p^2(Z_{d_L}+\Pi_{d_L})\right]\right\} \ ,
\label{eq-Vv}
\end{equation}
where the first term is the contribution from the top, the second term is the contribution from the Left-handed bottom, and the correlators are defined in Eq.~(\ref{eq-correlators}).

Expanding Eq.~(\ref{eq-Vv}) in powers of $1/Z_f$ (similar to an expansion in powers of $\epsilon_f$), to leading order we obtain:
\begin{equation}
V\simeq \alpha\ ({\rm c}_{v/f}-1)+\beta\ {\rm s}^2_{v/f}+\gamma\ {\rm s}^4_{v/f}+\delta\ {\rm c}_{v/f}\ {\rm s}^2_{v/f} +{\cal O}\left(\frac{1}{Z_f^2}\right)  \ ,
\label{eq-Vv1}
\end{equation}
with the coefficients given by:
\begin{align}
\alpha=-N_c\int\frac{d^4p}{(2\pi)^4}&\frac{\Pi_u^{(6,1,1,3)}-\Pi_u^{(6,1,3,1)}}{2Z_u} \ , \nonumber \\
\beta=-N_c\int\frac{d^4p}{(2\pi)^4}&\left[
\frac{8\Pi_q^{(6,3,1,1)}+5\Pi_q^{(6,1,1,3)}+5\Pi_q^{(6,1,3,1)}-18\Pi_q^{(6,3,2,2)}}{4Z_q}
%+3\frac{\Pi_{q'}^{(15,3,1,1)}-\Pi_{q'}^{(15,1,2,2)}}{2Z_{q'}}
\right. \nonumber\\
&\left.+\frac{3}{4Z_{u}}\left(2\Pi_{u}^{(6,3,2,2)}-\Pi_{u}^{(6,1,1,3)}-\Pi_{u}^{(6,1,3,1)}\right)
\right] \nonumber \\
\gamma=-N_c\int\frac{d^4p}{(2\pi)^4}&\left[
\frac{1}{Z_{q}}\left(4\Pi_{q}^{(6,3,2,2)}-2\Pi_{q}^{(6,3,1,1)}-\Pi_{q}^{(6,1,1,3)}-\Pi_{q}^{(6,1,3,1)}\right)
\right. \nonumber\\
&\left.+\frac{1}{4Z_{u}}\left(2\Pi_{u}^{(6,3,1,1)}+\Pi_{u}^{(6,1,1,3)}+\Pi_{u}^{(6,1,3,1)}-4\Pi_{u}^{(6,3,2,2)}\right)
\right] \nonumber \\
\delta=-N_c\int\frac{d^4p}{(2\pi)^4}&\left[\frac{\Pi_q^{(6,1,3,1)}-\Pi_q^{(6,1,1,3)}}{Z_q}+\frac{\Pi_u^{(6,1,1,3)}-\Pi_u^{(6,1,3,1)}}{2Z_u}\right] \ .
\end{align}

If the coefficients of the potential~(\ref{eq-Vv1}) are of the same order, the minimum of the potential is at $v=0$ or $v\sim f$. For ${\rm s}_{v/f}\ll 1$, the potential of Eq.~(\ref{eq-Vv1}) is minimised by:
\begin{equation}
{\rm s}_{v/f}^2\simeq\frac{-2\alpha-4(\beta+\delta)}{\alpha+8\gamma-4\delta} \ ,
\end{equation}
requiring tuning for a partial cancellation of the numerator. As usual the tuning is expected to be of order $\xi^{-1}$, see sec.~\ref{sec-tuning}.

Due to the presence of the trigonometric functions in the invariants, the potential of Eq.~(\ref{eq-Vv}) is invariant under the same transformations, and has the same periodicity with $v_4$, as Eq.~(\ref{eq-vSM}). As function of $v$, it has period $2\pi f$ .

\subsection{Numerical results}
In this section we present the results obtained by computing the potential of Eq.~(\ref{eq-Vv}). For numerical calculations, it is necessary to know the fermionic form-factors. An explicit realization can be obtained by working in a two-site model, with the elementary fields associated to the degrees of freedom of one site, and the first level of resonances of the SCFT associated to the degrees of freedom of the other site. We give a brief description of the sectors and show explicitly the form-factors in App.~\ref{ap-2site}. As mentioned before, since in our approach the potential is dominated by the third generation of quarks, we only include massive resonances associated to the doublet $q_L$ and the singlet $u$ of the third generation, both in the representation ${\cal R}={\bf 286}$ of SO(13). The masses of these multiplets of resonances, before mixing with the elementary sector and EWSB, are denoted as $M_q$ and $M_u$. Since both multiplets are in the same representation, an SO(13) invariant mass mixing term is allowed, whose coefficient we call $M_y$. By using the formalism of sec.~\ref{sec-fermions}, it is also possible to write Yukawa interactions between these fermionic resonances and the NGBs, we call these couplings $y_{\bf R}$, they are of order $g_*$. 

Below we describe a benchmark point of the parameter space, where the top and Higgs masses, as well as $v_{\rm SM}$, can be reproduced:
\begin{align}\label{eq-benchmark}
&\epsilon_{q3}=0.76\ , && \epsilon_{u3}=0.97\ , && f=1.63\ {\rm TeV}\ , \nonumber\\
&y_{({\bf 6},{\bf 3},{\bf 1},{\bf 1})}=-0.8\ ,&& y_{({\bf 6},{\bf 1},{\bf 3},{\bf 1})}=-1.6\ , && y_{({\bf 6},{\bf 1},{\bf 1},{\bf 3})}\sim-y_{({\bf 6},{\bf 2},{\bf 2},{\bf 3})}\sim 1 \ , \nonumber\\
&M_y=4.6\ {\rm TeV}\ , && M_q=2.3\ {\rm TeV}\ , && M_u=1.7\ {\rm TeV} \ ,
\end{align}
where $\tan\theta_{\psi}=\lambda_\psi f/M_\psi\equiv\epsilon_\psi$, with $\theta_\psi$ the angle diagonalising the mixing between the elementary fermions and the resonances. The values of $y_{({\bf 6},{\bf 1},{\bf 1},{\bf 3})}$ and $y_{({\bf 6},{\bf 3},{\bf 2},{\bf 2})}$ are, either allowed to vary in an interval, or fixed to values of ${\cal O}(1)$, we specify their values for each analysis done below. The other Yukawa couplings do not play any role in the minimization of the potential, as long as Eq.~(\ref{eq-min}) is satisfied, in the following sections they will be needed to determine, for example, the masses of the leptoquarks, we will give their values in those sections.  

\begin{figure}[t]
\centering
\includegraphics[width=0.47\textwidth]{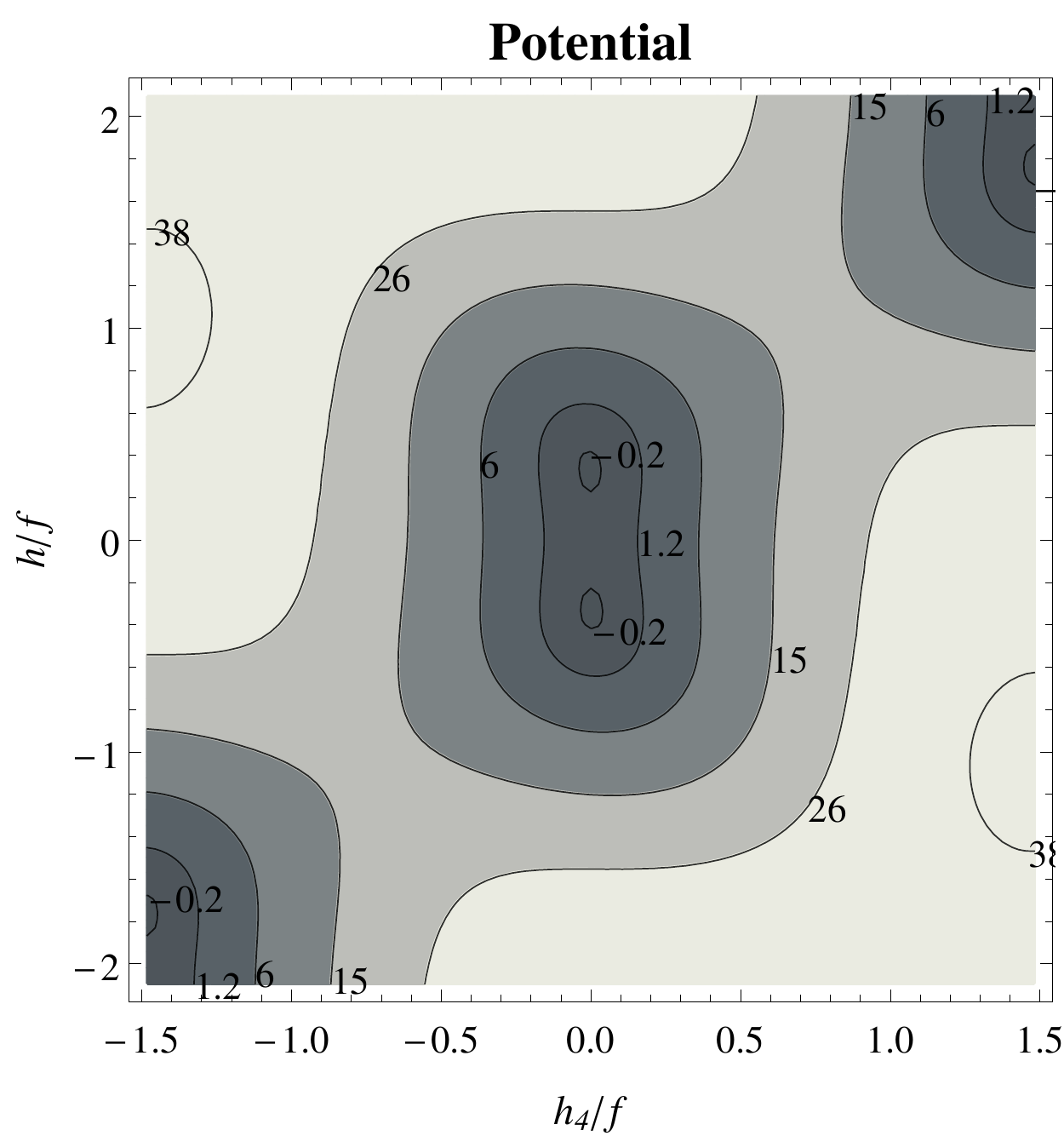}\hskip0.8cm
\includegraphics[width=0.47\textwidth]{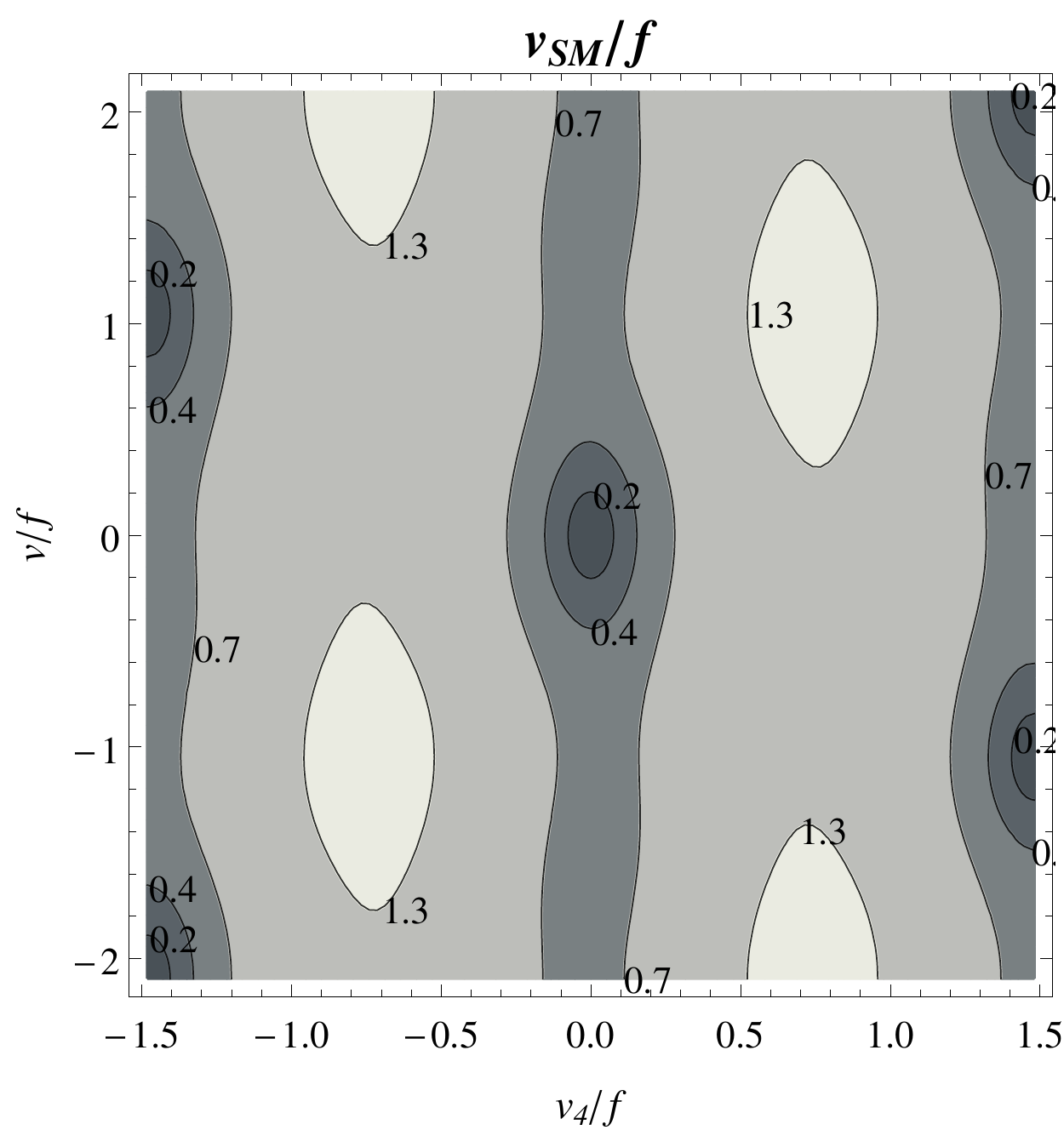}
\caption{On the left we show a contour plot of the potential, lighter (darker) gray shows higher (lower) values of the potential, and the labels on the contours indicate the height of the potential. The small contours with label -0.2 contain the minima of the potential defining $v$ and $v_4$. The parameters corresponding to this potential are defined by the benchmark point of Eq.~(\ref{eq-benchmark}) and $y_{({\bf 6},{\bf 1},{\bf 1},{\bf 3})}=1.1$ and $y_{({\bf 6},{\bf 3},{\bf 2},{\bf 2})}=-0.73$. On the right we show contour lines of $v_{\rm SM}/f$ defined in Eq.~(\ref{eq-vSM}). }
\label{fig-potential}
\end{figure}

As an example of the form of the potential, in the left panel of Fig.~\ref{fig-potential} we show $V$ as function of $h/f$ and $h_4/f$ for the benchmark point, with $y_{({\bf 6},{\bf 1},{\bf 1},{\bf 3})}=1.1$ and $y_{({\bf 6},{\bf 3},{\bf 2},{\bf 2})}=-0.73$. The lines indicate the height of the potential, with lighter gray for the maxima and darker gray for the minima, located inside the closed-curves with label -0.2. The plot exhibits the symmetries of the potential. 

Once the potential is minimised, fixing the value of $v$ and $v_4$, on the right panel of Fig.~\ref{fig-potential} one can read the value of $v_{\rm SM}/f$.

\begin{figure}[h]
\centering
\includegraphics[width=0.87\textwidth]{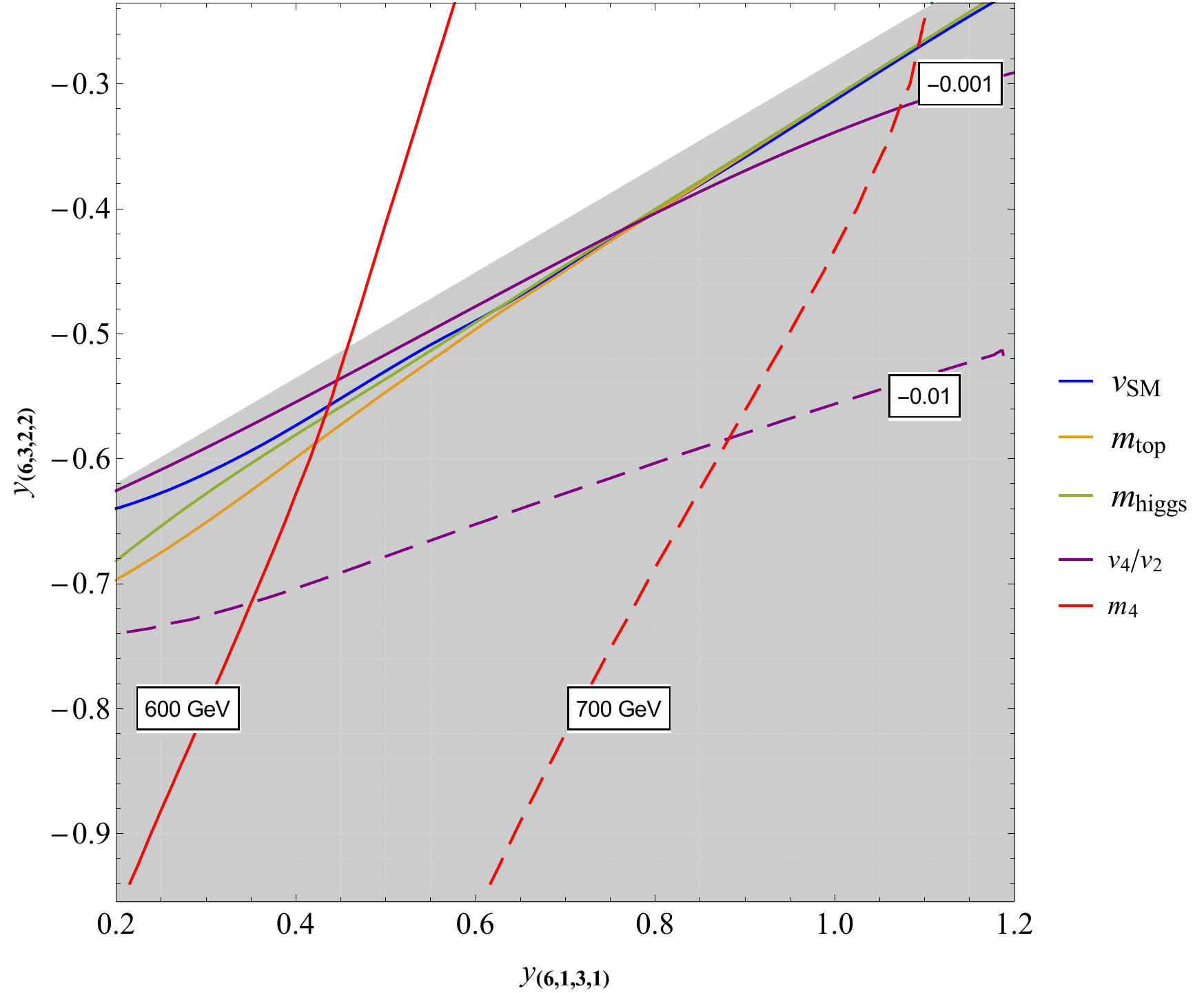}
\caption{In gray and white we show the regions with and without EWSB in the plane $y_{({\bf 6},{\bf 1},{\bf 1},{\bf 3})}-y_{({\bf 6},{\bf 3},{\bf 2},{\bf 2})}$, with the other parameters fixed by the benchmark point described in Eq.~(\ref{eq-benchmark}). In the plot $\xi$ increases from zero in the white region, to $\sim 0.41$ in the down-right corner, with $v_{\rm SM}=246$~GeV along the blue line. In the orange and green lines $m_t\simeq 150$~GeV and $m_h\simeq 125$~GeV, respectively. The violet lines indicate constant values of $v_4/v$, whereas the red ones show constant values of $m_{H4}$.}
\label{fig-planoplot}
\end{figure}

In Fig.~\ref{fig-planoplot} we show several interesting predictions of the model for the benchmark point, with $y_{({\bf 6},{\bf 1},{\bf 1},{\bf 3})}\in(0.2,1.2)$ and $y_{({\bf 6},{\bf 3},{\bf 2},{\bf 2})}\in(-0.95,-0.25)$. In the white region there is no EWSB, whereas in the gray area $v>0$. The blue line shows the region where $v_{\rm SM}$ takes the value of the SM, whereas the orange and green lines correspond to the regions where the top and the Higgs have masses: $m_t\simeq 150$~GeV and $m_h\simeq 125$~GeV. Around the region $y_{({\bf 6},{\bf 1},{\bf 1},{\bf 3})}=1$ and $y_{({\bf 6},{\bf 3},{\bf 2},{\bf 2})}=0.33$, $v_{\rm SM}$, $m_t$ and $m_h$ take simultaneously the values of the SM. The red lines show regions where $m_{H4}$, defined in Eq.~(\ref{eq-V42}), has constant values. Up to effects of EWSB, these values give the mass of the components of $H_4$. The violet lines show constant regions for $v_4/v$, as can be seen it is ${\cal O}(10^{-3})$ in the region that looks like the SM, and it becomes ${\cal O}(10^{-2})$ below that region. As explained above, the masses of the leptoquarks are not shown in this plot, because they depend on a set of Yukawa couplings that have not been fixed yet, we discuss them in sec.~\ref{sec-pheno}.

\subsection{Tuning}\label{sec-tuning} 
As is well known, EW precision tests require a separation between $v$ and $f$. Since generically the potential leads to no EWSB: $\xi=0$, or maximal EWSB: $\xi\sim 1$, an amount of tuning of order $\xi^{-1}$ is needed to obtain a separation between these scales. In composite Higgs models with custodial symmetry, EW precision tests require $\xi\lesssim 0.1-0.3$, the bound being mainly dominated by the $S$ parameter and $Zb_L\bar b_L$. In our model, as discussed in sec.~\ref{sec-constraints}, $v_4$ breaks the custodial symmetry, and the fermion embedding chosen does not protect the $Zb_L\bar b_L$ coupling, therefore we expect more tuning, compared with composite Higgs models with custodial protection of $g_{b_L}$, to pass the EWPT.

We use the sensitivity parameter defined in Refs.~\cite{Barbieri:1987fn,Anderson:1994dz,Panico:2012uw}, as an estimate of the fine-tuning of the model. We study the dependence of the potential on the parameters of the theory: the masses of the fermionic resonances, the composite Yukawa couplings, the mass mixing the fermionic resonances and the decay constant of the NGBs, as well as the degree of compositeness of the light fermions: $\epsilon_f$. 

For the benchmark of Eq.~(\ref{eq-benchmark}) the estimated tuning is $\sim\xi^{-1}\simeq 40$.
Calculating the tuning over the curve with $v_{SM} = 246$~GeV of fig. \ref{fig-planoplot}, we find the tuning to vary between 40 and 90, diminishing as $y_{({\bf 6,3,1,3 })}$ increases. If we explore higher values of this Yukawa, we find that the tuning can get as low as 25, when $y_{({\bf 6,3,1,3 })} \sim 1.7$, and after this increasing up to 200 as $y_{({\bf 6,3,1,3 })} \sim 5$. It is dominated by $\epsilon_q$ and $\epsilon_u$. 

\section{Phenomenology}\label{sec-pheno}
In this section we compute the corrections to $Zb_L\bar b_L$ induced at tree level by the presence of the fermionic resonances, showing that they saturate the bounds for the benchmark region of the parameter space. We also discuss some properties of the pNGBs interesting for their phenomenology, as their masses and couplings. Finally, we analyse the effect of the leptoquarks on flavor physics, as the $B$-anomalies, $B\to K^{(*)\nu\bar\nu}$ and lepton flavor universality violation, and we briefly comment on constraints from colliders.

\subsection{Corrections to $Z$ couplings}
As discussed in sec.~\ref{sec-constraints}, the composite fermions mixing with the elementary $b_L$ are not in the proper representation of the custodial symmetry to protect $g_{b_L}$. Describing the composite fermionic resonances with the two-site model defined in App.~\ref{ap-2site}, we have computed $\delta g_{b_L}$ at tree level by the following procedure. We have considered only the multiplets associated to $q$ and $u$ of the third generation, there are ten down-type fermions in representation ${\bf 286}$. With this content of fermionic resonances, we have computed the mass matrix of the down-sector, in the elementary-composite basis. We have performed a diagonalization of the mass matrix expanding in powers of $\xi$, in fact only the lightest eigenstate is needed for the calculation of $\delta g_{b_L}$. Expressing the interactions with the $Z$ in terms of the mass basis states, we obtained:
\begin{align}\label{eq-deltagbL}
\frac{\delta g_{b_L}}{g/c_W}\simeq& \xi \lambda_{q3}^2 f^4\left(24\{M_q^2M_u^2+f^2\lambda_{q3}^2[M_u^2+(M_y+fy_{({\bf 6},{\bf 3},{\bf 2},{\bf 2})})^2]\}\right)^{-1} \nonumber\\
&\left[
8y_{({\bf 6},{\bf 3},{\bf 1},{\bf 1})}^2
+y_{({\bf 6},{\bf 1},{\bf 3},{\bf 1})}^2
-16 y_{({\bf 6},{\bf 3},{\bf 1},{\bf 1})}y_{({\bf 6},{\bf 3},{\bf 2},{\bf 2})}
-2y_{({\bf 6},{\bf 1},{\bf 3},{\bf 1})}y_{({\bf 6},{\bf 3},{\bf 2},{\bf 2})}\right. \nonumber\\
&\left.+9
(y_{({\bf 6},{\bf 1},{\bf 1},{\bf 3})}^2
-2y_{({\bf 6},{\bf 1},{\bf 1},{\bf 3})}y_{({\bf 6},{\bf 3},{\bf 2},{\bf 2})}
+2y_{({\bf 6},{\bf 3},{\bf 2},{\bf 2})}^2
)
\right] +{\cal O}(\xi^2)\ .
\end{align}

The full mass matrix depends on a set of Yukawa couplings that have not been fixed in the benchmark point. For numerical results we have varied these couplings randomly, with $|y_{\bf R}|\in(0.3,\pi)$.
By comparison with the results obtained doing the full numerical diagonalization, we have verified that, for the region of the parameter space of Fig.~\ref{fig-planoplot}, the accuracy of Eq.~(\ref{eq-deltagbL}) is of percent level. For the region of Fig.~\ref{fig-planoplot} where $v_{\rm SM}$ is around the SM value, $\delta g_{b_L}\simeq (0.2- 0.4)\%$, with the smallest value for smaller $y_{({\bf 6},{\bf 1},{\bf 1},{\bf 3})}$, and increasing smoothly with this Yukawa. 

In sec.~\ref{sec-constraints} we estimated, up to factors of ${\cal O}(1)$ that depend on the representations ${\cal R}_q$ and ${\cal R}_u$, $\delta g_{b_L}\sim \xi\epsilon_q^2$. For the benchmark point this leads to $\delta g_{b_L}\sim 1\%$. Doing the calculation, we obtain that the Clebsch-Gordan coefficients, as well as the moderate values of the composite Yukawa, lead to an extra factor of order $0.2- 0.4$. Therefore, for the benchmark region of Fig.~\ref{fig-planoplot}, $\delta g_{b_L}$ is of the order of the bound from precision measurements.

As we will discuss in sec.~\ref{sec-banomalies}, it is also interesting to consider the possibility of large degree of compositeness of $\tau_L$. Eq.~(\ref{eq-frep13-3}) shows that the lepton doublets are embedded in a $({\bf 2},{\bf 2})_0$ of SU(2)$_L\times$SU(2)$_R\times$U(1)$_X$, thus the $Z\tau_L\bar \tau_L$ coupling is protected by $P_{LR}$ symmetry, allowing large $\epsilon_{\ell 3}$. This is not the case of $W$-interactions, as will be discussed in \ref{sec-FV}. Besides it is not possible to protect $Z\nu\bar\nu$ simultaneously with $Z\tau_L\bar \tau_L$, thus we expect corrections in the $Z$ coupling to neutrinos of the third generation, that will have an effect in the invisible width of the $Z$.

\subsection{Masses of the pNGBs}
Let us now discuss the spectrum of the pNGBs.
For the minimum of Eq.~(\ref{eq-min}), the masses of the leptoquarks and $H_4$ are estimated by the equation on the left of (\ref{eq-eV4}), in terms of the mass of a usual resonance: $m_\Phi\sim m_*(\epsilon_f g_*/4\pi)$, with $\Phi\neq H$. For $m_*\sim 2-10$ TeV, $\epsilon\sim 1$ and moderate values of $g_*\sim 2-5$, we expect: $m_\Phi\sim 0.4-3$ TeV.

After EWSB, the pNGBs with the same electric charge are mixed. Labelling the mass matrices with an index that indicates the electric charge of the states, for the color neutral scalars we obtain:
\begin{align}
&M^2_0= \left(
\begin{array}{ccc}
m_H^2+v^2\ 3 c_1 & -v^2\frac{3}{4}c_6 & 0 \\
\dots & m_{H_4}^2+v^2\left(\frac{c_4}{2}-\frac{c_5}{4\sqrt{10}}- \frac{2 c_8}{\sqrt{10}}\right)  & 0 \\
\dots & \dots & m_{H_4}^2+v^2\left(\frac{c_4}{2}-\frac{c_5}{4\sqrt{10}}+ \frac{2 c_8}{\sqrt{10}}\right)
\end{array}
\right) \ ,
\nonumber \\
&
M^2_1=\left(
\begin{array}{cc}
m_{H_4}^2+\frac{v^2}{4}\left(2 c_4+c_5\frac{1}{\sqrt{10}}\right) & v^2 \sqrt{\frac{3}{10}}c_8 \\
\dots & m_{H_4}^2+\frac{v^2}{4}\left(2 c_4-c_5\frac{3}{\sqrt{10}}\right)
\end{array}
\right)
\ , 
\\
&M^2_2=m_{H_4}^2+v^2\left(\frac12 c_4+c_5\frac{3}{4\sqrt{10}}\right)\ ,
\label{eq-mass-H} 
\end{align}
whereas the leptoquark mass matrices are given by:
\begin{align}
&M^2_{2/3}=
\left(
\begin{array}{cc}
m_{\tilde R_2}^2+v^2\frac{\sqrt{2}c_{30}+c_{31}}{4} & -v\frac{m_1}{\sqrt{3}}\\
\dots & m_{S_3}^2-v^2\frac{2\sqrt{3}c_{34}-3c_{35}}{12} 
\end{array}
\right) \ ,
\nonumber\\
&M^2_{-1/3}=
\left(
\begin{array}{ccc}
  m_{\tilde R_2}^2+v^2\frac{\sqrt{2}c_{30}-c_{31}}{4} & v\frac{-m_1}{\sqrt{6}} & -\frac{v^2}{2}c_{47}\\
\dots & m_{S_3}^2-v^2\frac{c_{34}}{2 \sqrt{3}} & v\frac{m_2}{2 \sqrt{3}}\\
\dots & \dots & m_{\hat R_2}^2+v^2\frac{\sqrt{2}c_{32}+c_{33}}{4} 
\end{array}
\right) \ ,
\nonumber\\
&M^2_{-4/3}= 
\left(
\begin{array}{cc}
m_{S_3}^2+v^2\frac{2\sqrt{3}c_{34}+3c_{35}}{12} & v\frac{m_2}{\sqrt{6}}\\
\dots & m_{\hat R_2}^2+v^2\frac{\sqrt{2}c_{32}-c_{33}}{4} 
\end{array}
\right) \ .
\label{eq-mass-LQ}
\end{align}
Since the mass matrices are symmetric, we have not written the elements of the left-down block. From the diagonal elements of the mass matrices it is straightforward to identify the basis, as an example, for leptoquarks $S^{-4/3}$, the basis is: $\{S_3^{-4/3},\hat R_2^{-4/3}\}$, whereas for the colorless neutral states, the basis is: $\{H, {\rm Re}[H_4^{(0)}],{\rm Im}[H_4^{(0)}] \}$. The coefficients $m_i$  are the cubic couplings of the potential, Eq.~(\ref{eq-V43}), whereas $c_i$ are the quartic ones, Eq.~(\ref{eq-V44}), thus they can be written in terms of integrals of the correlators, as detailed in App.~\ref{ap-quartic}. Besides their size they can be estimated by using Eq.~(\ref{eq-eV4}).

Let us consider first an analytical study of the spectrum of scalars, and after that we present some numerical results.
For the analysis of the spectrum of the neutral states, we trade $m_H^2\to -c_1v^2$ in $M_0^2$, as required from the minimization of the potential, Eq.~(\ref{eq-vmin}), and we diagonalise $M_0^2$. The lightest neutral state, to be identified with the physical Higgs, has a mass: $m_0^2\simeq 2c_1v^2$, with corrections suppressed by powers of $\xi$. This state is to leading order given by the neutral component of the doublet $H$. To next order it mixes with the neutral states in $H_4$, with mixing angle $\frac{3c_6v^2}{4m_{H_4}^2}\sim\xi$. The other neutral states receive corrections from the Higgs vev: $m_{1,2}^2\simeq m_{H_4}^2+v^2\left(c_4 \frac12-c_5\frac{1}{4\sqrt{10}}\mp c_8\frac{2}{\sqrt{10}}\right)$, that induces a splitting between them. We have checked this approximation in the numeric analysis of the one-loop potential, performed to all orders in $\xi$.

The masses of the charged states also receive corrections from the Higgs vev. The splitting of the states with charge +1 is of order $v^2\sqrt{\left(\frac{3}{16} c_5^2+ c_8^2\right)\frac{6}{5}}$.

The masses of the leptoquarks are corrected by the Higgs vev also, that induces splittings $\delta m^2_{LQ}\sim {\cal O}(c_jv^2)$. Since the non-diagonal terms of the mass matrices are of order $v$, instead of $v^2$ as for the colorless states, the mixing angles of the leptoquarks are ${\cal O}(\sqrt{\xi})$.

For a numerical study of the masses, we define two separate regions of the parameter space in terms of the benchmark region of Eq.~(\ref{eq-benchmark}) and the following Yukawa couplings:

\begin{tabular}{|c|c|c|c|c|c|c|c|}
\hline\rule{0mm}{5mm}
region & $y_{({\bf 6, 3, 2, 2})}$ & $y_{({\bf 15, 3, 1, 1})}$ & $y_{({\bf 15, 1, 2, 2})}$ & $y_{({\bf 1, 1, 2, 2})}$ & $y_{({ \bf 1, 3, 2, 2})}$ & $y_{({\bf 1, 3, 3, 1 })}$ & $y_{({\bf 1, 3, 1, 3 })}$
\\[5pt]
\hline \rule{0mm}{5mm}
A & -1.51 & -0.58 & -1.08 & -0.15 & -1.36 & -0.79 & 1.38
\\
\hline\rule{0mm}{5mm}
B & 1.51 & -0.63 & -1.36 & -0.72 & -1.13 & -1.23 & -1.41
\\
\hline 
\end{tabular}

\begin{figure}[h]
  \includegraphics[width=\textwidth]{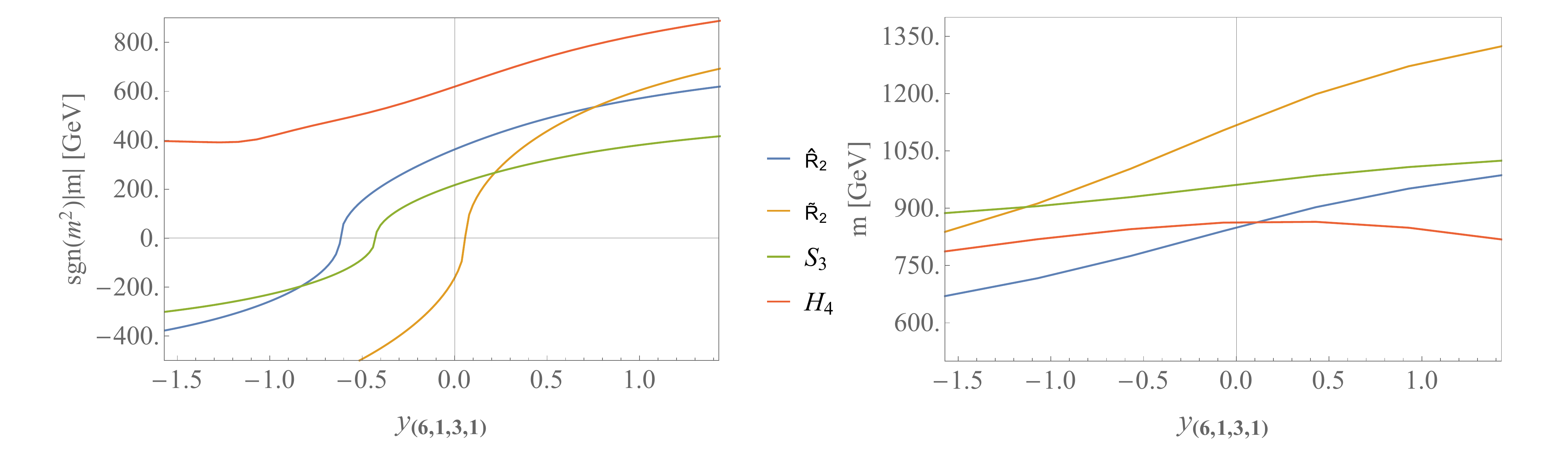}
      \centering
      \caption{Plot of leptoquark masses along with $m_4$, for two different regimes. On the left frame, in a region where the masses squared become negative. We plot the absolute value of the mass, along with the sign of $m^2$. On the right frame we plot a region where leptoquark masses are all higher, reaching about a TeV.}
\label{masslq}
\end{figure}

We show our results in Fig.~\ref{masslq}, region A on the left and region B on the right, we do not take into account the effect of the Higgs vev in those plots. For region A, the masses can vary quite abruptly, from an order TeV to vanishing values. It is also possible to obtain negative squared masses, although in this case the quadratic coefficient is not the mass, and there can be breaking of $\textrm{SU}(3)_c$. On the right we show a typical region where the masses acquire larger values, with positive squares. 

We have also computed the effect of $v$. In region B there is no splitting since there is no EWSB, $v = 0$. In region A we calculated the splittings between components of each leptoquark multiplet. These splittings (with respect to the masses before EWSB) get as large as $\sim 20\%$ near $y_{({ \bf 6,1,3,1 } )} \sim -1.5$, where the masses squared are negative. As this Yukawa increases and the masses become real, their splittings become lower. For $S_3$ they are lower than $1 \%$, for $\su$, around $2 \%$, and for $\sd$ around $3 \%$.

By a random scan over all the Yukawa couplings in the interval $[-\pi/2,\pi/2]$, we find that the dominant Yukawas are $y_{({ \bf 6,1,3,1 } )}$ and $y_{({\bf 6,3,2,2})}$.

\subsection{Couplings of the leptoquarks}
Other very important quantity for the phenomenology of the pNGBs, is their coupling with the SM fields.
Expanding Eq.~(\ref{eq-Leff1}) in powers of the NGBs it is possible to obtain the Yukawa interactions with the fields $H$, $H_4$ and the leptoquarks. The flavor structure of the couplings is determined by the structure of the mixings $\epsilon_f$, as well as by the anarchic structure of the SCFT. They can be estimated as:~\cite{Agashe:2004cp}
\begin{equation}
y_{ff'}\sim c_{ff'}\ g_*\ \epsilon_f\epsilon_{f'} \ ,
\label{eq-yff}
\end{equation}
with the dimensionless factor $c_{ff'}\sim {\cal O}(1)$.

Expanding to first order in the leptoquarks and to second order in $H$, we obtain the following leptoquark interactions:
\begin{align}
{\cal L}_{\rm int}\supset &  
%H(y_u\bar q^u u+y_d\bar q^d d+y_e\bar \ell e)+
y_3 S_3 \bar{q^c_L} \ell_L+\frac{1}{f}H (y_{3,1}S_3 \bar{q_L} e_R+y_{2,1}\su\bar{q^c_L} \ell_L+y_{3,2}S_3\bar{d_R} \ell_L) \nonumber
\\
&+\frac{1}{f^2}H^2(y_{3,3}S_3 \bar{q^c_L} \ell_L+y_{2,2}\su \bar{q_L} e_R+y_{3,4}S_3\bar{d^c_R} e_R+y_{2,3}y \sd \bar{d_R} \ell_L) + {\rm h.c.}
\label{eq-yLQ}
\end{align}
where flavor indices are understood.
Due to the structure of the unbroken group H and the embedding of the quarks and leptons, only $S_3$ interacts with operators of dimension four. Interactions with $\hat R_2$ and $\tilde R_2$ are only present at the level of higher dimensional operators involving the Higgs. For this reason their effect in the phenomenology is suppressed compared with $S_3$, in particular the impact of $\tilde R_2$ in $R_K^{(*)}$ can be neglected.

The couplings of Eq.~(\ref{eq-yLQ}) can be expressed in terms of the fermionic correlators. A good approximation can be obtained by evaluating the correlators at zero momentum. We get:
\be\label{eq-y3}
y_3 = \frac{\Pi_{q,\ell}^{(1,3,2,2)}(0)}{\sqrt{Z_{\ell} + \Pi_{\ell,\ell}^{(1,3,2,2)}(0)}\sqrt{Z_q + \Pi_{q,q}^{(6,3,2,2)}(0)}}
\te 
The fact that $y_3$ depends on $\Pi_{q,\ell}^{(1,3,2,2)}$ only, can be understood from the following simple argument. The only way of contracting the dressed fields, when evaluating Eq.~(\ref{eq-Leff1}) at first order in the NGB fields, is by choosing either ${\bf R}_{\ell} = {\bf (1,3,2,2)} $, or ${\bf R}_q = {\bf (6,3,2,2)}$. However, only ${\bf{R}}_\ell$ is among the common H-representations in the decomposition of ${\bf 78}$ and ${\bf 286}$. The denominator of Eq.~(\ref{eq-y3}) arises after canonical normalization of the fermion fields.

The couplings of Eq.~(\ref{eq-yLQ}) are not expected to be aligned in flavor space. This happens because different correlators depend on different combinations of composite Yukawa couplings that, having uncorrelated flavor structures, lead to couplings with the SM fields that are not aligned. A full numerical calculation of them would require the introduction of three generations of composite resonances, as well as elementary fermions. We have not done that calculation in the present work, instead we will use the estimates of Eq.~(\ref{eq-yff}) in the following. 

\subsection{Analysis of flavor physics}
We analyse some of the most important effects of the leptoquarks on flavor transitions and lepton flavor universality violation. We do not perform a full analysis, instead we discuss their effect in the $B$-anomalies, as well as the largest constraints. Since the interactions with $\tilde R_2$ and $\hat R_2$ are suppressed by positive powers of $\xi$, we only consider the effects from $S_3$. In the following we will make extensive use of the bounds presented in Ref.~\cite{Buttazzo:2017ixm}.

At low energies the leptoquarks can be integrated out at tree-level, leading to the following effective Lagrangian
\begin{equation}\label{eq-l4f}
{\cal L}_{\rm eff}\supset\frac{C}{v^2}[(\bar q_L\gamma^\mu\sigma^aq_L)(\bar \ell_L\gamma_\mu\sigma^a\ell_L)+3(\bar q_L\gamma^\mu q_L)(\bar \ell_L\gamma_\mu\ell_L)] \ ,
\end{equation}
where generation indices are understood. The dimensionless coefficient $C$ is given by~\footnote{For comparison with the literature: $C_T=-C^{3333}$ and $C_S=-3C^{3333}$, minus the coefficients of the current-triplet and -singlet when all the fermions are in the third generation.}

\begin{equation}\label{eq-C3}
C^{ijkl}=y_{3,il}y_{3,jk}^*\frac{v^2}{4m_{S_3}^2}\sim c_{il}c_{jk}^* \epsilon_{qi}\epsilon_{qj}\epsilon_{\ell k}\epsilon_{\ell l}\frac{g_*^2v^2}{4m_{S_3}^2} \ .
\end{equation}

\subsubsection{B-anomalies}\label{sec-banomalies}
It is well known that for suitable values of the leptoquark Yukawa couplings, an $S_3$ at the TeV scale can explain the deviations in $R_K$ and $R_{K^{*}}$. Following Ref.~\cite{Buttazzo:2017ixm}, a global fit of $b\to s\mu\mu$ (neglecting effects in $ee$) gives: 
\begin{equation}\label{eq-b1}
\Delta C_9^\mu=-\Delta C_{10}^\mu=\frac{4\pi}{\alpha_{\rm em}V_{tb}V_{ts}^*}C^{2322}=-0.61\pm 0.12 \ .
\end{equation}

By making use of Eqs.~(\ref{eq-mixq}), (\ref{eq-mixl}), (\ref{eq-eV4}) and (\ref{eq-b1}), in our model we obtain:
\begin{equation}
g_*^{3/2}f\sim 4\ {\rm TeV} \ ,
\label{eq-b2}
\end{equation}
up to factors of ${\cal O}(1)$. This equation fits nicely with $f\sim$ TeV and moderate values of $g_*$.

Let us comment now on the anomalies on $R_{D^{(*)}}\equiv R_{b\to c}^{\tau \ell}$. A fit to deviations from $\tau$ universality in $b\to c\ell\bar\nu$ gives:~\cite{Buttazzo:2017ixm}
\begin{equation}\label{eq-b3}
R_{D^{(*)}}\simeq 1+2 \sum_j C^{3j33}\frac{V_{cj}}{V_{cb}}=1.237\pm 0.053 \ .
\end{equation}

Again we make use of Eqs.~(\ref{eq-mixq}), (\ref{eq-mixl}), (\ref{eq-eV4}) and (\ref{eq-b3}), but this time we keep the dependence on $\epsilon_{l3}$. Eq.~(\ref{eq-b3}) requires
\begin{equation}
\frac{g_*f}{\epsilon_{\ell 3}}\sim 10\ {\rm TeV} \ ,
\label{eq-b4}
\end{equation}

From Eqs.~(\ref{eq-b2}) and (\ref{eq-b4}), we obtain that, in order to simultaneously explain $R_{K^{(*)}}$ and $R_{D^{(*)}}$: $\sqrt{g_*}\epsilon_{\ell 3}\sim 0.4$. One must compare this result with the estimate for the $\tau$-Higgs Yukawa coupling in the case of $\epsilon_{\ell i}=\epsilon_{e i}$, Eq.~(\ref{eq-mixl}), that gives $\sqrt{g_*}\epsilon_{\ell 3}\sim 0.08$. Thus, in order to explain $R_{D^{(*)}}$, one has to abandon the assumption of similar degree of compositeness of both chiralities of the $\tau$, and consider instead the case $\epsilon_{e 3}/ \epsilon_{\ell 3}\sim 0.04$. In this case, although $Z\tau _L\bar \tau _L$ is protected, large $\epsilon_{\ell 3}$ induces corrections in the $W$ couplings with the $\tau$ lepton.~\cite{Buttazzo:2017ixm} In the next section we will show that the bounds from precision measurement of this coupling do not allow to fit $R_{D^{(*)}}$.

\subsubsection{Constraints from $\tau$ decays and $B\to K^{(*)}\nu\bar\nu$}\label{sec-FV}
In the present scenario the tightest bounds in flavor physics arise from flavor universality violation in $\tau$ decays. $B\to K^{(*)}\nu\bar\nu$ is also a good process to look for effects of the leptoquarks, since neutrinos of third generation can potentially give large contributions. We will not perform a full analysis of flavor observables, instead we will analyse these two processes in the presence of the low energy effective interactions of Eq.~(\ref{eq-l4f}).

One-loop corrections to the $W$ coupling in the presence of leptoquarks give:\cite{Feruglio:2016gvd,Buttazzo:2017ixm}
\begin{equation}\label{eq-Wtau1}
\left|\frac{g_{W\tau}}{g_{W\ell}}\right|=1-\frac{6y_t^2}{(4\pi)^2}C^{3333}\log\frac{\Lambda}{m_t}\simeq 1-0.084C^{3333} \ ,
\end{equation}
where $\Lambda$ has been fixed to 2 TeV. Departures of this coupling from lepton flavor universality can not be large than per mil level. Making use of Eqs.~(\ref{eq-mixq}), (\ref{eq-mixl}), (\ref{eq-eV4}) and (\ref{eq-Wtau1}), and leaving the dependence on $\epsilon_{\ell 3}$, we obtain:
\begin{equation}
\frac{g_*f}{\epsilon_{\ell 3}}\gtrsim 28\ {\rm TeV} \ .
\label{eq-Wtau2}
\end{equation}
Using Eq.~(\ref{eq-b2}) in (\ref{eq-Wtau2}), we obtain: $\sqrt{g_*}\epsilon_{\ell 3}\lesssim 0.15$, that can be easily satisfied for $\epsilon_{\ell 3}\simeq\epsilon_{e 3}$, but is smaller than the value needed to fit $R_{D^{(*)}}$.

The 95\%CL bound on $B\to K^{(*)}\nu\bar\nu$~\cite{Patrignani:2016xqp,Buttazzo:2017ixm} in our model can be approximated by:
\begin{equation}\label{eq-BKnn1}
B_{K^{(*)}\nu\bar\nu}\simeq\frac{1}{3}\left[2+\left|1+\frac{2\pi}{\alpha_{\rm em}}\frac{C^{3233}}{C^{\rm SM}_\nu V^*_{ts}V_{tb}}\right|^2\right]<5.2 \ ,
\end{equation}
where $B_{K^{(*)}\nu\bar\nu}\equiv {\cal B}(B\to K^{(*)}\nu\bar\nu)_{\rm exp}/{\cal B}(B\to K^{(*)}\nu\bar\nu)_{\rm SM}$ and $C^{\rm SM}_\nu=-6.4$. Making use of Eqs.~(\ref{eq-mixq}), (\ref{eq-mixl}), (\ref{eq-eV4}) and (\ref{eq-BKnn1}), and keeping $\epsilon_{\ell 3}$, we obtain:
\begin{equation}
\frac{g_*f}{\epsilon_{\ell 3}}\gtrsim 17- 22\ {\rm TeV} \ ,
\label{eq-BKnn2}
\end{equation}
depending on the complex phase of the correction. Using Eq.~(\ref{eq-b2}) in (\ref{eq-BKnn2}), we obtain: $\sqrt{g_*}\epsilon_{\ell 3}\lesssim 0.2- 0.25$, again compatible with $\epsilon_{\ell 3}\simeq\epsilon_{e 3}$, but smaller than the value needed to fit $R_{D^{(*)}}$.

\subsection{Collider physics}
We discuss very briefly constraints of our model at colliders, and we mention some interesting signals.

Direct searches of new physics also give constraints on the leptoquarks, the most important ones from pair production by QCD interactions at LHC. Different analysis of the collected data give bounds on scalar leptoquark masses that are roughly of order 1 TeV~\cite{Dorsner:2016wpm,Dorsner:2018ynv,Sirunyan:2018nkj,Alvarez:2018gxs}, to be compared with the predictions for these masses in the present model, that are $\sim 0.4- 1.2$~TeV, for $f\sim 1.6$~TeV. 

Other production processes are: single production, that has been studied, for example, in Refs.~\cite{Buttazzo:2017ixm,Dorsner:2018ynv}, and non-resonant production, that can be found in~\cite{Alvarez:2018jfb}. Single and non-resonant leptoquark production at LHC are more model dependent, since they depend on the leptoquark Yukawa couplings with the SM fermions, that are not fixed. The framework of partial compositeness gives an estimate of the size of these couplings. These processes, with couplings compatible with partial compositeness, become competitive for leptoquarks masses larger than $1- 1.5$ TeV.

In the present case, with larger couplings to quarks and leptons of the third generation, one can expect interesting phenomena associated with top and bottom quark production, as well as tau leptons. Final states with muon leptons are also interesting, due to the cleaner final state. In all the cases, promising channels are those with multi-leptons. We refer the reader to the references of the previous paragraphs of this section, and references therein, for detailed analysis of the collider phenomenology. 

Another very interesting signal at colliders is the creation of fermionic resonances. As usual in models with custodial symmetry and partial compositeness, there are custodians with masses that can be lighter than $m_*$, as well as exotic charges, for example quark partners with $Q=5/3$ and $-4/3$. The composite grand unified symmetry also leads to quarks with exotics representations under SU(3)$_c$ (se App.~\ref{ap-SO(13)}), as color octets and sextets, that could be created in pairs by QCD interactions. For the benchmark region of the parameter space, the lightest fermionic resonances have masses of order 1 TeV. 

The fermionic resonances will decay to SM particles, or to SM particles and leptoquarks. Particularly interesting is the decay of the color octets and sextets. Color octets arising from the multiplet ${\bf 286}$ can be doublets or triplets of SU(2)$_L$. They will decay to a leptoquark plus a quark, preferentially of the third generation: $\Psi^{\bf 8}\to qLQ\to qq'\ell$, with $q$ and $q'$ being dominantly top and bottom quarks, and $\ell$ being a tau lepton. Color sextets would decay to a leptoquark plus an octet resonance, posibly off-shell, with the following decay of the octet as described in the previous sentence: $\Psi^{\bf 6}\to\Psi^{\bf 8(*)}LQ\to qLQLQ\to qq'q''\ell\ell'$. 

Summarizing, a very reach phenomenology involving leptoquarks and exotic fermions is expected, with preferential decays to third generation SM fermions. A detailed study of their production and detection is beyond the scope of this work.

\section{Discussions and conclusions}\label{sec-conclusions}
The $B$-anomalies are one of the most exciting phenomena reported by experiments in the last years. Leptoquarks at the TeV scale could be responsible for them. In the present work we have given an effective description of a new strongly interacting sector at the TeV scale, that contains leptoquarks and Higgses as NGBs. The global symmetry group was chosen as the minimal simple group containing the SM plus the custodial symmetry, and able to deliver the Higgs and a leptoquark $S_3$ as NGBs.
Given the pattern of global symmetry breaking, the content of leptoquarks and Higgses was fixed, in our case, besides the Higgs, a colorless SU(2)$_L$-fourplet and three leptoquarks were present: an $S_3\sim(\bar{\bf 3},{\bf 3})_{1/3}$, as well as two EW doublets transforming as $({\bf 3},{\bf 2})_{1/6}$ and $({\bf 3},{\bf 2})_{-5/6}$. The assumption of anarchic partial compositeness of the SM fermions, as well as the choice of the representations of the fermionic resonances under the global symmetry, determined the structure of Yukawa couplings and the structure of the potential. We have shown that the interactions with the SM fermions can trigger EWSB successfully, and generate leptoquarks masses of order TeV. By modelling the resonances of the SCFT with a two-site theory, we have computed the one-loop potential and the spectrum of pNGBs. We have found a benchmark region of the parameter space where the masses of the SM states: the $W$, the top and Higgs, are around their experimental values, and the pNGBs have masses of order $0.4-1.3$~TeV, with a NGB decay constant $f=1.6$~TeV.

Some amount of tuning is needed to obtain a separation between the EW scale and the NGB decay constant, that characterises the scale of the SCFT. We found that, for the benchmark region analysed, the tuning is dominated by the degree of compositeness of the quarks of the third generation, varying between 40 and 90 for $v_{\rm SM}\simeq 246$~GeV. Those values are compatible with the estimate given by $\xi^{-1}\simeq 40$. 

We have analysed several constraints, as the corrections to the $\rho$ parameter due to the vev of the colorless fourplet, and the $Z$-couplings. We have shown that the vev of $H_4$ is suppressed by $\xi$, in agreement with the results of Ref.~\cite{Babu:2009aq}. For the benchmark region of the parameter space where $v_{\rm SM}\simeq 246$~GeV we obtained: $v_4/v\sim{\cal O}(10^{-3})$, allowing to pass constraints from the $\rho$ parameter. For the $Z$-couplings, since the resonances that mix with $b_L$ were embedded in a representation of the custodial symmetry that does not allow to protect $Zb_L\bar b_L$, there could be large corrections. For the benchmark region of the parameter space we obtained $\delta g_{b_L}/g\simeq 0.2- 0.4\%$, saturating the bound from precision measurements. Tighter bounds would require, either a larger tuning, or a larger representation of the fermionic resonances, allowing custodial protection of $g_{b_L}$, as in the case of the SO(13) representation ${\cal R}_q=\overline{\bf 715}$. Thus within the present model, and for $f\sim 1.3- 2$~TeV, deviations of order few per mil can be expected.

As discussed in Refs.~\cite{Crivellin:2017zlb,Buttazzo:2017ixm,Marzocca:2018wcf}, the presence of an additional leptoquark in the representation $(\bar{\bf 3},{\bf 1})_{1/3}$, with a mass similar to that of $S_3$ and couplings with the same flavor structure, would allow to explain simultaneously $R_{K^{(*)}}$ and $R_{D^{(*)}}$, without too large corrections to flavor processes (as violation of lepton flavor universality in $W$ coupling to $\tau$, or in flavor changing neutral current decays as $B\to K\nu\bar\nu$). Also a new leptoquark in $({\bf 3},{\bf 2})_{7/6}$ could be a possibility.~\cite{Becirevic:2018afm}
It is straightforward to include states with these charges in the present model, but not as NGBs, instead they would be ordinary resonances, with larger masses. In this case, it is not possible in general to pass bounds from flavor physics (fine tuning would be needed to ensure, for example, a partial cancellation of the Wilson coefficients of dangerous operators). An interesting possibility would be to find a simple group able to generate these states also as NGBs, as well as embeddings of the SM fermions leading to the right Yukawa couplings. 

Composite grand unified models also predict the presence of exotic states, like color octets and sextets, with cascade decays to quarks and leptons of the third generation. The study of their phenomenology at colliders certainly deserves a dedicated analysis.

\section*{Acknowledgements}
We thank Ezequiel \'Alvarez and Aurelio Juste for discussions, and ICAS-UNSAM for hospitality in several stages of this project. This work was partially supported by the Argentinian ANPCyT PICT 2013-2266.

%%%%%%%%%%%%%%%%%%%%%%%%%%%%%%%%%%%%%%%%%%%%%%%%%%%%%%%%%%%%%%%%

\appendix
\section{Representations of SO(13)}\label{ap-SO(13)}

In this appendix we give a brief description of the algebra, as well as the lowest dimensional representations, of the group SO(3).
A simple basis for the algebra of SO(13) in the fundamental representation is given by the set of generators $\{T_{\ell m},\ell<m=2,\dots 13\}$, with coefficients:
\begin{equation}
({\cal T}_{\ell, m})_{jk}=i(\delta_{\ell j}\delta_{mk}-\delta_{mj}\delta_{\ell k}) \ , \qquad l<m \ .
\end{equation}

An SO(7)$\times$SO(6) subgroup can be defined by the transformations leaving invariant the block diagonal matrix:
\begin{equation}
A=\left(\begin{array}{cc}a I_7 & 0\\0 & b I_6\end{array}\right) \ ,
\end{equation}
with $a$ and $b$ different non-trivial numbers, $I_n$ the identity matrix in $n$-dimensions and the action of the group being defined as:
\begin{equation}
A\rightarrow U A U^\dagger , \qquad U\in SO(13) \ .
\end{equation}
An algebra of SU(2)$_a\times$SU(2)$_b\times$SU(2)$_c$ inside SO(7) can be defined by:
\begin{align}
&T^a_1=-\frac{1}{2}({\cal T}_{1,4}+{\cal T}_{2,3}) \ ,&
&T^a_2=\frac{1}{2}({\cal T}_{1,3}-{\cal T}_{2,4}) \ , &
&T^a_3=-\frac{1}{2}({\cal T}_{1,2}+{\cal T}_{3,4}) \ , \nonumber
\\
&T^b_1=\frac{1}{2}({\cal T}_{1,4}-{\cal T}_{2,3}) \ ,&
&T^b_2=\frac{1}{2}({\cal T}_{1,3}+{\cal T}_{2,4}) \ , &
&T^b_3=-\frac{1}{2}({\cal T}_{1,2}-{\cal T}_{3,4}) \ , \nonumber
\\
&T^c_1={\cal T}_{5,6} \ ,&
&T^c_2={\cal T}_{5,7} \ ,&
&T^c_3={\cal T}_{6,7} \ ,
\label{alg-su23}
\end{align}
An algebra of SU(3)$\times$U(1) inside SO(6) can be defined by:
\begin{align}
&T^{\rm SU(3)}_1=\frac{1}{2}({\cal T}_{10,13}-{\cal T}_{11,12}) \ ,&
&T^{\rm SU(3)}_2=\frac{1}{2}({\cal T}_{10,12}+{\cal T}_{11,13}) \ ,\nonumber \\
&T^{\rm SU(3)}_3=\frac{1}{2}(-{\cal T}_{10,11}+{\cal T}_{12,13}) \ ,&
&T^{\rm SU(3)}_4=\frac{1}{2}({\cal T}_{8,13}-{\cal T}_{9,12}) \ ,\nonumber \\
&T^{\rm SU(3)}_5=\frac{1}{2}({\cal T}_{8,12}+{\cal T}_{9,13}) \ ,&
&T^{\rm SU(3)}_6=\frac{1}{2}({\cal T}_{8,11}-{\cal T}_{9,10}) \ ,\nonumber \\
&T^{\rm SU(3)}_7=\frac{1}{2}({\cal T}_{8,10}+{\cal T}_{9,11}) \ ,&
&T^{\rm SU(3)}_8=\frac{1}{2\sqrt{3}}(-2{\cal T}_{8,9}+{\cal T}_{10,11}+{\cal T}_{12,13}) \ ,\nonumber\\
&T^{\rm U(1)}=-4({\cal T}_{8,9}+{\cal T}_{10,11}+{\cal T}_{12,13}) \ , 
\label{alg-su31}
\end{align}

The parity $P=e^{iT_P\pi/2}$ can be obtained with $T_P=T^{\rm U(1)}$.

The adjoint representation ({\bf 78}), can be obtained by using the structure constants, or by using the algebra itself as a basis of the vector space of dimension 78.

The representation ${\bf 286}$ can be obtained, for example, from the product ${\bf 13}\otimes {\bf 78}\sim {\bf 13}\oplus {\bf 286} \oplus {\bf 715}$. Although we have built it explicitly for our calculations, we will not show the generators here because the matrices are too large.

The smallest representations of SO(13), and their decompositions under H are:
\begin{align}
& {\bf 13}\sim ({\bf 6},{\bf 1},{\bf 1},{\bf 1}) \oplus({\bf 1},{\bf 1},{\bf 2},{\bf 2}) \oplus({\bf 1},{\bf 3},{\bf 1},{\bf 1})\ ,  \nonumber
\\
& {\bf 64}\sim ({\bf 4},{\bf 1},{\bf 2},{\bf 2}) \oplus({\bf 4},{\bf 2},{\bf 1},{\bf 2}) \oplus{\rm c.c.}\ , \nonumber
\\
& {\bf 78}\sim ({\bf 15},{\bf 1},{\bf 1},{\bf 1}) \oplus({\bf 1},{\bf 3},{\bf 1},{\bf 1}) \oplus({\bf 1},{\bf 1},{\bf 3},{\bf 1}) \oplus({\bf 1},{\bf 1},{\bf 1},{\bf 3}) \oplus({\bf 1},{\bf 3},{\bf 2},{\bf 2}) \oplus({\bf 6},{\bf 3},{\bf 1},{\bf 1}) \nonumber
\\
&\qquad \oplus({\bf 6},{\bf 1},{\bf 2},{\bf 2})\ , \nonumber
\\
& {\bf 286}\sim ({\bf 15},{\bf 3},{\bf 1},{\bf 1}) \oplus({\bf 15},{\bf 1},{\bf 2},{\bf 2}) \oplus({\bf 1},{\bf 1},{\bf 1},{\bf 1}) \oplus({\bf 1},{\bf 1},{\bf 2},{\bf 2}) \oplus({\bf 1},{\bf 3},{\bf 2},{\bf 2}) \oplus({\bf 1},{\bf 3},{\bf 1},{\bf 3}) \nonumber 
\\
&\quad \oplus({\bf 1},{\bf 3},{\bf 3},{\bf 1}) \oplus({\bf 10},{\bf 1},{\bf 1},{\bf 1}) \oplus({\bf 6},{\bf 3},{\bf 2},{\bf 2}) \oplus({\bf 6},{\bf 3},{\bf 1},{\bf 1}) \oplus({\bf 6},{\bf 1},{\bf 3},{\bf 1}) \oplus({\bf 6},{\bf 1},{\bf 1},{\bf 3}) \oplus{\rm c.c.}
\label{eq-frep13-1}
\end{align}
the complex conjugate representations must be added only when they are not equivalent to the original one.

These representations can be further decomposed under H$_{\rm min}$ to see which of them contain the SM fermions. Using the decompositions of Eq.~(\ref{eq-so6-su3}) and the identification ${\rm SU(2)_{L}}\equiv{\rm SU(2)_{1+2}}$, we obtain:
\begin{align}
& {\bf 13}\sim ({\bf 3},{\bf 1},{\bf 1})_{2}\oplus({\bf 1},{\bf 2},{\bf 2})_0\oplus({\bf 1},{\bf 3},{\bf 1})_0\oplus{\rm c.c.}\ , \nonumber
\\
& {\bf 64}\sim ({\bf 3},{\bf 2},{\bf 2})_{-1}\oplus({\bf 3},{\bf 1},{\bf 1})_{-1}\oplus({\bf 3},{\bf 3},{\bf 1})_{-1}\oplus({\bf 1},{\bf 2},{\bf 2})_3\oplus({\bf 1},{\bf 3},{\bf 1})_3\oplus({\bf 1},{\bf 1},{\bf 1})_3\oplus{\rm c.c.}\ , \nonumber
\\
& {\bf 78}\sim({\bf 8},{\bf 1},{\bf 1})_0\oplus({\bf 3},{\bf 1},{\bf 1})_{-4}\oplus({\bf 3},{\bf 3},{\bf 1})_2\oplus({\bf 3},{\bf 2},{\bf 2})_{2}\nonumber
\\
&\qquad \oplus({\bf 1},{\bf 4},{\bf 2})_0\oplus({\bf 1},{\bf 2},{\bf 2})_0\oplus({\bf 1},{\bf 1},{\bf 3})_0\oplus 2({\bf 1},{\bf 3},{\bf 1})_0\oplus({\bf 1},{\bf 1},{\bf 1})_0\oplus{\rm c.c.} \ , \nonumber
\\
%& 286\sim ({\bf 1},{\bf 1},{\bf 1})_6\oplus({\bf 1},{\bf 3},{\bf 1})_0\oplus({\bf 8},{\bf 3},{\bf 1})_0\oplus({\bf 8},{\bf 2},{\bf 2})_0\oplus({\bf 1},{\bf 2},{\bf 2})_0 \nonumber 
%\\
%&\qquad \oplus({\bf 3},{\bf 1},{\bf 1})_2\oplus({\bf 6},{\bf 1},{\bf 1})_{-2}\oplus({\bf 3},{\bf 3},{\bf 1})_{-4}\oplus({\bf 3},{\bf 2},{\bf 2})_{-4}\oplus({\bf 3},{\bf 3},{\bf 1})_2\oplus({\bf 3},{\bf 1},{\bf 3})_2\oplus({\bf 3},{\bf 2},{\bf 2})_2\oplus({\bf 3},{\bf 4},{\bf 2})_2 \oplus{\rm c.c.} \ ,
%\\
%& 286\sim ({\bf 3},{\bf 1},{\bf 1})_2\oplus({\bf 6},{\bf 1},{\bf 1})_{-2}\oplus({\bf 3},{\bf 3},{\bf 1})_{-4}\oplus({\bf 3},{\bf 2},{\bf 2})_{-4}\oplus({\bf 3},{\bf 3},{\bf 1})_2\oplus({\bf 3},{\bf 1},{\bf 3})_2\oplus({\bf 3},{\bf 2},{\bf 2})_2\oplus({\bf 3},{\bf 4},{\bf 2})_2 \oplus{\rm c.c.} \nonumber 
%\\
%&\qquad\oplus({\bf 8},{\bf 3},{\bf 1})_0\oplus({\bf 8},{\bf 2},{\bf 2})_0 \oplus({\bf 1},{\bf 1},{\bf 1})_6\oplus({\bf 1},{\bf 3},{\bf 1})_0\oplus({\bf 1},{\bf 2},{\bf 2})_0 \ ,
%\\
& {\bf 286}\sim ({\bf 8},{\bf 3},{\bf 1})_0\oplus({\bf 8},{\bf 2},{\bf 2})_0\oplus({\bf 6},{\bf 1},{\bf 1})_{-2}\oplus({\bf 3},{\bf 3},{\bf 1})_{-4}\oplus 2({\bf 3},{\bf 3},{\bf 1})_2\oplus({\bf 3},{\bf 4},{\bf 2})_2 \nonumber
\\
&\qquad \oplus({\bf 3},{\bf 2},{\bf 2})_{-4}\oplus({\bf 3},{\bf 2},{\bf 2})_2\oplus({\bf 3},{\bf 1},{\bf 3})_2\oplus({\bf 3},{\bf 1},{\bf 1})_2\oplus ({\bf 1},{\bf 3},{\bf 3})_{0}\oplus ({\bf 1},{\bf 4},{\bf 2})_{0} \nonumber
\\
&\qquad \oplus ({\bf 1},{\bf 5},{\bf 1})_{0}\oplus 3({\bf 1},{\bf 2},{\bf 2})_0\oplus 2({\bf 1},{\bf 3},{\bf 1})_0\oplus({\bf 1},{\bf 1},{\bf 1})_6\oplus 2({\bf 1},{\bf 1},{\bf 1})_{0}\oplus{\rm c.c.}\ . 
\label{eq-frep13-2}
\end{align}

From Eq.~(\ref{eq-frep13-2}) it is straightforward to see that, besides the representations of Eq.~(\ref{eq-frep13-3}), there are other embeddings containing the SM fermions, for example $d_R$ could be embedded into $({\bf 6},{\bf 1},{\bf 1},{\bf 1})\subset {\bf 13}$ and $\ell_L$ into $({\bf 1},{\bf 1},{\bf 2},{\bf 2})\subset {\bf 13}$. However, the H symmetry does not allow the proper dimension-four Yukawa couplings with $H$ and $S_3$ for these embeddings.

\section{Potential}\label{ap-potential}

To obtain a series expansion of the potential in powers of the NGBs, we expand the matrix ${\cal K}$, defined in Eq.~(\ref{eq-VCW}), as a series in powers of $\Pi$. We add a spurion $\epsilon$, to be fixed to $\epsilon=1$ in the end of the calculation, and trade $\Pi\to\epsilon\Pi$. After this replacement ${\cal K}$ can be written as
\begin{equation} 
{\cal K}=\sum_{n=0}^\infty \epsilon^n {\cal K}_n\ .
\end{equation} 
We subtract from the potential a constant divergent term, independent of $\Pi$, given by the momentum integral of $\log\det{\cal K}_0$. Using the operator identity: $\log\det{\cal K}={\rm tr}\log{\cal K}$, we obtain the following expression for the integrand of $V$, that we call $w$:
\begin{equation} 
w\equiv-({\rm tr}\log{\cal K}-{\rm tr}\log{\cal K}_0)=- {\rm tr}\log({\cal K}_0^{-1}{\cal K})=- {\rm tr}\log( 1 + \epsilon {\cal K}_0^{-1} {\cal K}_1 + \dots) \ .
\end{equation}
the minus sign is from the fermionic loop. By Taylor expanding to ${\cal O}(\epsilon^4)$ we get
\begin{align} 
-w\simeq& \;  \epsilon \, {\rm tr}\left( \widetilde{\cal K}_1 \right)  + \epsilon^2 \, {\rm tr}\left( \widetilde{\cal K}_2  - \frac{\widetilde{\cal K}_1^2}{2} \right)  + \epsilon^3 \, {\rm tr}\left(\widetilde{\cal K}_3 -  \widetilde{\cal K}_1 \widetilde{\cal K}_2 + \frac13 \widetilde{\cal K}_1^3     \right)  
\nonumber \\
&+ \epsilon^4 {\rm tr}\left(\widetilde{\cal K}_4 - \widetilde{\cal K}_1 \widetilde{\cal K}_3 - \frac12 \widetilde{\cal K}_2^2 + 12 \widetilde{\cal K}_1^2 \widetilde{\cal K}_2 - \frac14 \widetilde{\cal K}_1^4  \right) + {\cal O}(\epsilon^5)
\end{align}
with the definition $\widetilde{\cal K}_k \equiv {\cal K}_0^{-1} {\cal K}_k$.

This expansion allows to obtain the potential to fourth order in the NGBs, it is straightforward to go to higher orders. In performing the expansion, we have traded the complicated calculation of the determinant of a rather large matrix, by the much simpler calculation of traces.

By keeping the effect of ${\bf R}_{q'}$, defined below Eq.~(\ref{eq-frep13-3}), and choosing a basis for the fermions: $\{f\}=\{u_R^i,u_L^i,d_L^i,{u'}_L^i,{d'}_L^i\}$, with $i=1,2,3$ being a color index, the ${\cal K}_n$ matrices can be computed. For example: ${\cal K}_0$ is a diagonal matrix that can be written as
\begin{equation} 
{\cal K}_0={\rm diag}(Z_u+\Pi_{uu}^{(6,1,1,3)}I_3,Z_q+\Pi_{qq}^{(6,3,2,2)}I_6,Z_{q'}+\Pi_{q'q'}^{(15,1,2,2)}I_6) \ ,
\end{equation}
where the first block is for the d.o.f. associated to the three colors of $u$, and the other two blocks account for the d.o.f. associated to $q$ and $q'$. Notice that this matrix is of dimension 15, if the effect of ${\bf R}_{q'}$ is neglected, by taking its mixing very small, the resulting ${\cal K}$ is of dimension~9.

\section{Invariants of quartic potential}\label{ap-quartic}
In this appendix we show the quartic terms of the potential.

The quartic order contains 49 singlets, of which 8 are composed only of $H$ and $H_4$, 20 only of leptoquarks, and the remaining 21 of $H$ and leptoquarks. All these singlets were built using Clebsch-Gordan coefficients for SU(2), and for SU(3) the following product rules 
\begin{align} 
\mathbf{3} \times \mathbf{\bar{3}} &\sim \mathbf{1} + \mathbf{8}  \nonumber  \\ 
\mathbf{3} \times \mathbf{3} &\sim \mathbf{\bar{3}}_A + \mathbf{6}_S \nonumber 
\end{align} 
Where the $A$ and $S$ subscripts stand for anti-symmetric and symmetric products, respectively. For the $\mathbf{8}$ representation, we used the Gell-Mann matrices $\lambda^a$, $a \in \lbrace 1, .. ,8 \rbrace $. For instance, if $\psi$ and $\phi$ are two fields transforming in the $\mathbf{3}$ of SU(3), if we form the products
\begin{equation} 
\sum_{ij} \; \lambda^a_{ij}\,  \bar{\phi}_i \, \psi_j \equiv  O^a
\end{equation} 
Then this object $O^a$ transforms in the $ \mathbf{8}$ (octet) representation.

We make a list of linear independent operators, using the following notation: when making the product of two representations, we will denote with a subindex in what representation of $SU(3)\times SU(2)$ it transforms, or when dealing only with color singlets, just the $SU(2)$ representation. Just with fields $H$ and $H_4$:
\begin{align}
&{\cal Q}_1=\left((H H^\dg)_{(1)}\right)^2 ,&
& {\cal Q}_2=(H_4H_4)_{(3)} \cdot (H_4^\dg H_4^\dg)_{(3)} , \nonumber\\
& {\cal Q}_3=(H_4H_4)_{(7)} \cdot (H_4^\dg H_4^\dg)_{(7)} ,&
& {\cal Q}_4=(HH^\dg)_{(1)}\cdot (H_4H_4^\dg)_{(1)} , \nonumber\\
& {\cal Q}_5=(HH^\dg)_{(3)}\cdot (H_4H_4^\dg)_{(3)} ,&
& {\cal Q}_6=(H^\dg H^\dg)_{(3)}\cdot (H H_4)_{(3)} + \hc , \nonumber\\
& {\cal Q}_7=(H_4^\dg H_4^\dg)_{(3)}\cdot (H H_4)_{(3)} + \hc ,&
& {\cal Q}_8=(H H)_{(3)}\cdot (H_4^\dg H_4^\dg)_{(3)} + \hc .
\end{align}
Purely leptoquarks:
\al
& {\cal Q}_9=(S_3 S_3)_{(6,1)}\cdot (S_3^\dg S_3^\dg)_{(\bar{6},1)} \ ,&
& {\cal Q}_{10}=(S_3 S_3)_{(6,5)}\cdot (S_3^\dg S_3^\dg)_{(\bar{6},5)}\ ,\nonumber\\
& {\cal Q}_{11}=(S_3 S_3)_{(\bar{3},3)}\cdot (S_3^\dg S_3^\dg)_{(3,3)}\ ,&
& {\cal Q}_{12}=(\su \su)_{(\bar{3},1)}\cdot(\suh \suh)_{(3,1)}\ ,\nonumber\\
& {\cal Q}_{13}=(\su \su)_{(6,3)}\cdot(\suh \suh)_{(\bar{6},3)}\ ,&
& {\cal Q}_{14}=(\sd \sd)_{(\bar{3},1)}\cdot(\sdh \sdh)_{(3,1)}\ ,\nonumber\\
& {\cal Q}_{15}=(\sd \sd)_{(6,3)}\cdot(\sdh \sdh)_{(\bar{6},3)}\ ,&
& {\cal Q}_{16}=(\su \suh)_{(1,1)} \cdot (\sd \sdh)_{(1,1)}\ ,\nonumber\\
& {\cal Q}_{17}=(\su \suh)_{(1,3)} \cdot (\sd \sdh)_{(1,3)}\ ,&
& {\cal Q}_{18}=(\su \suh)_{(8,1)} \cdot (\sd \sdh)_{(8,1)}\ ,\nonumber\\
& {\cal Q}_{19}=(\su \suh)_{(8,3)} \cdot (\sd \sdh)_{(8,3)}\ ,&
& {\cal Q}_{20}=(\su \suh)_{(1,1)} \cdot (S_3 S_3^\dg)_{(1,1)}\ ,\nonumber\\
& {\cal Q}_{21}=(\su \suh)_{(1,3)} \cdot (S_3 S_3^\dg)_{(1,3)}\ ,&
& {\cal Q}_{22}=(\su \suh)_{(8,1)} \cdot (S_3 S_3^\dg)_{(8,1)}\ ,\nonumber\\
& {\cal Q}_{23}=(\su \suh)_{(8,3)} \cdot (S_3 S_3^\dg)_{(8,3)}\ ,&
& {\cal Q}_{24}=(\sd \sdh)_{(1,1)} \cdot (S_3 S_3^\dg)_{(1,1)}\ ,\nonumber\\
& {\cal Q}_{25}=(\sd \sdh)_{(1,3)} \cdot (S_3 S_3^\dg)_{(1,3)}\ ,&
& {\cal Q}_{26}=(\sd \sdh)_{(8,1)} \cdot (S_3 S_3^\dg)_{(8,1)}\ ,\nonumber\\
& {\cal Q}_{27}=(\sd \sdh)_{(8,3)} \cdot (S_3 S_3^\dg)_{(8,3)}\ ,&
& {\cal Q}_{28}=(S_3 S_3)_{(3,3)} \cdot (\su \sd)_{(\bar{3},3)} + \hc\ ,\nonumber\\
& {\cal Q}_{29}=(S_3 S_3)_{(\bar{6},1)} \cdot (\su \sd)_{(6,1)} + \hc\ . & &
\fal
The operators with two leptoquarks and two color singlets are:
\al
&{\cal Q}_{30}=(H H^\dg)_{(1)} (\su \suh)_{(1,1)}\ ,& 
&{\cal Q}_{31}=(H H^\dg)_{(3)} \cdot (\su \suh)_{(1,3)}\ ,\nonumber\\
&{\cal Q}_{32}=(H H^\dg)_{(1)} (\sd \sdh)_{(1,1)}\ ,& 
&{\cal Q}_{33}=(H H^\dg)_{(3)} \cdot (\sd \sdh)_{(1,3)}\ ,\nonumber\\
&{\cal Q}_{34}=(H H^\dg)_{(1)} (S_3 S_3^\dg)_{(1,1)}\ ,&
&{\cal Q}_{35}=(H H^\dg)_{(3)} \cdot (S_3 S_3^\dg)_{(1,3)}\ , \nonumber\\
&{\cal Q}_{36}=(H_4 H_4^\dg)_{(1)} (\su \suh)_{(1,1)}\ ,&
&{\cal Q}_{37}=(H_4 H_4^\dg)_{(3)} \cdot (\su \suh)_{(1,3)}\ ,\nonumber\\
&{\cal Q}_{38}=(H_4 H_4^\dg)_{(1)} (\sd \sdh)_{(1,1)}\ ,&
&{\cal Q}_{39}=(H_4 H_4^\dg)_{(3)} \cdot (\sd \sdh)_{(1,3)}\ ,\nonumber\\  
&{\cal Q}_{40}=(H_4 H_4^\dg)_{(1)} (S_3 S_3^\dg)_{(1,1)}\ ,&  
&{\cal Q}_{41}=(H_4 H_4^\dg)_{(3)} \cdot (S_3 S_3^\dg)_{(1,3)}\ ,\nonumber\\
&{\cal Q}_{42}=(H_4 H_4^\dg)_{(5)} \cdot (S_3 S_3^\dg)_{(1,5)}\ ,&  
&{\cal Q}_{43}=(H H_4^\dg)_{(3)} \cdot (\su \suh)_{(1,3)} + \hc\ ,\nonumber\\
&{\cal Q}_{44}=(H H_4^\dg)_{(3)} \cdot (\sd \sdh)_{(1,3)} + \hc\ ,& 
&{\cal Q}_{45}=(H H_4^\dg)_{(3)} \cdot (S_3 S_3^\dg)_{(1,3)} + \hc , \nonumber\\
&{\cal Q}_{46}=(H H_4^\dg)_{(5)} \cdot (S_3 S_3^\dg)_{(1,5)} + \hc\ ,& 
&{\cal Q}_{47}=(H H)_{(3)} \cdot (\sd \suh)_{(1,3)} + \hc , \nonumber\\
&{\cal Q}_{48}=(H_4 H_4)_{(3)} \cdot (\sd \suh)_{(1,3)} + \hc\ ,& 
&{\cal Q}_{49}=(H H_4)_{(3)} \cdot (\sd \suh)_{(1,3)} + \hc 
\fal

\section{Matching couplings}\label{ap-match}
In this appendix we give explicit expressions of the couplings of the effective theory in terms of the fermionic correlators and their momentum integrals. We keep the dependence on ${\bf R}_{q'}$.

The quadratic coefficients of the potential are:
\begin{align}
m_H^2=&-\int \frac{d^4p}{(2\pi)^4} \Bigg[ \frac14 \frac{24 \Pi_{q,q}^{(6,3,1,1)}+ 3 \Pi_{q,q}^{(6,1,1,3)}+ 27 (\Pi_{q,q}^{(6,3,2,2)}-2\Pi_{q,q}^{(6,1,3,1)})  }{Z_q + \Pi_{q,q}^{(6,3,2,2)}} \nonumber \\
    &     + \frac92 \frac{\Pi_{q',q'}^{(15,3,1,1)} - \Pi_{q',q'}^{(15,1,2,2)}}{Z_{q'} + \Pi_{q',q'}^{(15,1,2,2)}} + \frac92 \frac{ \Pi_{u,u}^{(6,3,2,2)}- \Pi_{u,u}^{(6,1,1,3)} }{Z_u + \Pi_{u,u}^{(6,1,1,3)}} - \frac92  \frac{ \left(\Pi_{q,u}^{(6,3,2,2)} - \Pi_{q,u}^{(6,1,1,3)}\right)^2 }{ (Z_q + \Pi_{q,q}^{(6,3,2,2)} )( Z_u + \Pi_{u,u}^{(6,1,1,3)}  ) } \Bigg]  \nonumber \\
m_{H_4}^2=&-\int\frac{d^4p}{(2\pi)^4}   \Bigg[ \frac32 \frac{ \Pi_{q,q}^{(6,3,1,1)}+ 2 \Pi_{q,q}^{(6,1,1,3)} -3 \Pi_{q,q}^{(6,3,2,2)}  }{Z_q + \Pi_{q,q}^{(6,3,2,2)}}     +  \nonumber
\\& \frac92 \frac{\Pi_{q',q'}^{(15,3,1,1)} - \Pi_{q',q'}^{(15,1,2,2)}}{Z_{q'} + \Pi_{q',q'}^{(15,1,2,2)}} + \frac92 \frac{ \Pi_{u,u}^{(6,3,2,2)}- \Pi_{u,u}^{(6,1,1,3)} }{Z_u + \Pi_{u,u}^{(6,1,1,3)}}  \Bigg] \nonumber
\\
m_{S_3}^2=&-\int\frac{d^4p}{(2\pi)^4}  \Bigg[ \frac{2 \Pi_{q,q}^{(1,3,2,2)} -7 \Pi_{q,q}^{(6,3,2,2)}+ 5 \Pi_{q,q}^{(15,1,2,2)}  }{Z_q + \Pi_{q,q}^{(6,3,2,2)}}  + 6 \frac{\Pi_{q',q'}^{(6,3,2,2)} - \Pi_{q',q'}^{(15,1,2,2)}}{Z_{q'} + \Pi_{q',q'}^{(15,1,2,2)}} \nonumber \\
    &     + \frac32 \frac{ \Pi_{u,u}^{(1,1,3,3)}- \Pi_{u,u}^{(6,1,1,3)} }{Z_u + \Pi_{u,u}^{(6,1,1,3)}} -  \frac{ \left(\Pi_{q',q^C}^{(6,3,2,2)} - \Pi_{q',q^C}^{(15,1,2,2)}\right)^2 }{ (Z_q + \Pi_{q,q}^{(6,3,2,2)} )( Z_{q'} + \Pi_{q',q'}^{(15,1,2,2)}  ) } \Bigg]  \nonumber 
\end{align}
\begin{align}
%\\ 
m_{\tilde R_2}^2=&-\int\frac{d^4p}{(2\pi)^4}  \Bigg[ \frac34 \frac{3 \Pi_{q,q}^{(1,3,1,3)} + \Pi_{q,q}^{(1,3,3,1)} -10 \Pi_{q,q}^{(6,3,2,2)} + 6 \Pi_{q,q}^{(15,3,1,1)} }{Z_q + \Pi_{q,q}^{(6,3,2,2)}} \nonumber \\
    &     + \frac34 \frac{ 6 \Pi_{q',q'}^{(6,1,1,3)}+ 2\Pi_{q',q'}^{(6,1,3,1)}+ 3\Pi_{q',q'}^{(10,1,1,1)}+3\Pi_{q',q'}^{(\bar{10},1,1,1)} - 14 \Pi_{q',q'}^{(15,1,2,2)}}{Z_{q'} + \Pi_{q',q'}^{(15,1,2,2)}} \nonumber
\\ &
+\frac32 \frac{ \Pi_{u,u}^{(1,1,2,2)}-4 \Pi_{u,u}^{(6,1,3,1)} + 3 \Pi_{u,u}^{(15,1,2,2)} }{Z_u + \Pi_{u,u}^{(6,1,1,3)}} \Bigg]  \nonumber \nonumber
\\
m_{\hat R_2}^2=&-\int\frac{d^4p}{(2\pi)^4}  \Bigg[ \frac32 \frac{ \Pi_{q,q}^{(1,3,3,1)} - 3  \Pi_{q,q}^{(6,3,2,2)}+ 2  \Pi_{q,q}^{(15,3,1,1)}  }{Z_q + \Pi_{q,q}^{(6,3,2,2)}}+ 3 \frac{ \Pi_{u,u}^{(15,1,2,2)}- \Pi_{u,u}^{(6,1,3,1)} }{Z_u + \Pi_{u,u}^{(6,1,1,3)}} \nonumber \\
    &     + \frac34 \frac{4 \Pi_{q',q'}^{(6,1,3,1)} + \Pi_{q',q'}^{(10,1,1,1)} + \Pi_{q',q'}^{(\bar{10},1,1,1)} - 6 \Pi_{q',q'}^{(15,1,2,2)} }{Z_{q'} + \Pi_{q',q'}^{(15,1,2,2)}}  -\frac32 \frac{ \left(\Pi_{q',u^C}^{(6,1,3,1)} - \Pi_{q',u^C}^{(15,1,2,2)}\right)^2 }{ (Z_u + \Pi_{u,u}^{(15,1,2,2)} )( Z_{q'} + \Pi_{q',q'}^{(15,1,2,2)}  ) } \Bigg] 
\end{align}

The coefficients $m_1$ and $m_2$ of Eq.~(\ref{eq-mass-LQ}) correspond to cubics involving $H$. We have, at leading order in $1/Z_f$,
\begin{align}
  m_2 &= - \int \frac{d^4p}{(2\pi)^4} \frac{3}{8 \sqrt6} \nonumber\\
&\Bigg[ \frac{ 8 \Pi_{q,q}^{(1,3,2,2)}-6\Pi_{q,q}^{(1,3,3,1)}-3\Pi_{q,q}^{(6,1,1,3)}-15\Pi_{q,q}^{(6,1,3,1)}+8\Pi_{q,q}^{(6,3,1,1)}+28\Pi_{q,q}^{(15,1,2,2)}-20\Pi_{q,q}^{(15,3,1,1)} }{Z_q} \nonumber \\
    &+ \frac{12\Pi_{q',q'}^{(6,1,3,1)}-3\Pi_{q',q'}^{(10,1,1,1)}-3\Pi_{q',q'}^{(\bar{10},1,1,1)}-6\Pi_{q',q'}^{(15,3,1,1)}}{Z_{q'}} + \frac{6\Pi_{u,u}^{(1,3,3,1)}+6\Pi_{u,u}^{(6,3,2,2)}-12 \Pi_{u,u}^{(15,1,2,2)}}{Z_u} \Bigg]
\end{align}
with a similar structure for $m_1$

For the quartics, the most of the coefficients are much longer, even for a $1/Z_f$ expansion. We can present, for example, some of the shortest:
\begin{align}
  c_6 &= - \int \frac{d^4p}{(2\pi)^4} \frac{3}{4 \sqrt2} \Bigg[ \frac{\Pi_{u,u}^{(6,1,1,3)}+\Pi_{u,u}^{(6,1,3,1)}+2\Pi_{u,u}^{(6,3,1,1)}-4 \Pi_{u,u}^{(6,3,2,2)}}{Z_u} \Bigg] \nonumber\\
  c_7 &= - \int \frac{d^4p}{(2\pi)^4} \frac{\sqrt5}{} \Bigg[ \frac12 \frac{3\Pi_{q,q}^{(6,3,2,2)}-\Pi_{q,q}^{(6,1,1,3)}-2\Pi_{q,q}^{(6,3,1,1)} }{Z_q} \nonumber \\
    &+ \frac32 \frac{\Pi_{q',q'}^{(15,3,1,1)}-\Pi_{q',q'}^{(15,1,2,2)}}{Z_{q'}} + \frac34 \frac{\Pi_{u,u}^{(6,3,1,1)}-\Pi_{u,u}^{(6,1,1,3)}}{Z_u} \Bigg]
\end{align}
Whereas other coefficients can get as much as 20 terms at first order in $1/Z_f$.

\section{Two-site theory}\label{ap-2site}
In this section we show the fermionic form factors that are obtained in a two-site theory. In this kind of theories the elementary sector is identified with one site, and the first level of resonances of the SCFT with another site. On the composite sector we include vector-like fermion resonances $\Psi^Q$ and $\Psi^U$, with masses $M_{q,u}\sim g_* f/\sqrt{2}$, of order few TeV. As described in sec.~\ref{sec-frep}, these fermions are in the representation {\bf 286} of SO(13). To obtain a finite one-loop potential we only include NGB interactions with the chiral structure $y_{\bf R}f(\bar \Psi^Q_LU)_{\bf R}(U^\dagger \Psi^U_R)_{\bf R}$, as well as a term $M_y\bar \Psi^Q_L\Psi^U_R$, see Ref.~\cite{Carena:2014ria}. Both sites interact through a $\sigma$-model field transforming bilinearly under the symmetries of both sites, with mixing $\lambda_q$ and $\lambda_u$, Eq.~(\ref{eq-pceff}). For a more detailed description we suggest the reading of Refs.~\cite{DeCurtis:2011yx,Contino:2006nn}. In the present case we follow the notation of Ref.~\cite {Andres:2015oqa}. 

The form factors are given by 
\begin{align}
& \Pi^{\bf R}_{f,f}(p)=\lambda_f^2 f^2\frac{M_f^2-p^2+y_{\bf R}^2f^2}{d_{\bf R}} \ ,
\qquad f=q,u \ ,\nonumber \\
& \Pi^{\bf R}_{q,u}(p)=-\lambda_q\lambda_u f^2\frac{M_qM_uy_{\bf R}f+M_y(p^2-y_{\bf R}^2f^2)}{d_{\bf R}} \ ,\nonumber \\
&d_{\bf R}=p^2(M_u^2+M_q^2)-M_q^2M_u^2+2M_qM_uM_yy_{\bf R}f+(M_y^2-p^2)(p^2-y_{\bf R}^2f^2) \ .
\end{align}

%\bibliography{biblio}{}
%\bibliographystyle{plain}

\bibliographystyle{JHEP}
\bibliography{biblio_v2}
\end{document}